\documentclass[twocolumn,twoside]{IEEEtran}

\ifCLASSINFOpdf

\else

\fi

\ifCLASSOPTIONcompsoc
    \usepackage[caption=false, font=normalsize, labelfont=sf, textfont=sf]{subfig}
\else
    \usepackage[caption=false, font=normalsize]{subfig}
\fi
\usepackage{lipsum}%
\usepackage[dvipsnames]{xcolor}
\usepackage{algorithm,algorithmic}
\usepackage{balance}
\usepackage{multicol}   
\usepackage{cite}
\usepackage{gensymb}
\usepackage{multirow}
\usepackage{graphics}
\usepackage{epsfig}
\usepackage{graphicx}
\usepackage{epstopdf}
\usepackage{textcomp}
\usepackage{amsmath}
\usepackage{mathtools}
\usepackage{filecontents}
\usepackage{lipsum,color}
\usepackage{amssymb}
\usepackage{float}
\usepackage{colortbl} 
\usepackage{times} 
\usepackage{amsthm}  

\usepackage{amsfonts}

\theoremstyle{break}
\begin{document}
\title{LoS Coverage Analysis for
	UAV-based THz Communication Networks: Towards 3D Visualization of Wireless Networks}

\author{ Mohammad~T.~Dabiri,~Mazen~Hasna,~{\it Senior Member,~IEEE},~Saud~Althunibat,~{\it Senior Member,~IEEE}, \\~and
	~Khalid Qaraqe,~{\it Senior Member,~IEEE}
\thanks{This publication was made possible by NPRP14C-0909-210008 from the Qatar National Research Fund (a member of The Qatar Foundation). The statements made herein are solely the responsibility of the author[s].}
\thanks{Mohammad Taghi Dabiri, and Mazen Hasna are with the Department of Electrical Engineering, Qatar University, Doha, Qatar.  (E-mail: m.dabiri@qu.edu.qa; hasna@qu.edu.qa).}
\thanks{Saud Althunibat is with the Department of Communications Engineering, Al-Hussein Bin Talal University, Jordan (E-mail: saud.althunibat@ahu.edu.jo).}
\thanks{Khalid A. Qaraqe is with the Department of Electrical and Computer Engineering, Texas A$\&$M University at Qatar, Doha 23874, Qatar (E-mail: khalid.qaraqe@qatar.tamu.edu).}
}

\maketitle
\begin{abstract}
Terahertz (THz) links require a line-of-sight (LoS) connection, which is hard to be obtained in most scenarios. For THz communications, analyses based on LoS probability are not accurate, and a new real LoS model should be considered to determine the LoS status of the link in a real 3D environment. Considering unmanned aerial vehicle (UAV)-based THz networks, LoS coverage is analyzed in this work, where nodes are accurately determined to be in LoS or not. 
Specifically, by modeling an environment with 3D blocks, our target is to locate a set of UAVs equipped with directional THz links to provide LoS connectivity for the distributed users among the 3D obstacles.
To this end, we first characterize and model the environment with 3D blocks. Then, we propose a user-friendly algorithm based on 3D spatial vectors, which determines the LoS status of nodes in the target area. In addition, using 3D modeling, several meta-heuristic algorithms are proposed for UAVs' positioning under 3D blocks in order to maximize the LoS coverage percentage. In the second part of the paper, for a UAV-based THz communication network, a geometrical analysis-based algorithm is proposed, which jointly clusters the distributed nodes and locates the set of UAVs to maximize average network capacity while ensuring the LoS state of distributed nodes among 3D obstacles. Moreover, we also propose a sub-optimal hybrid k-means-geometrical-based algorithm with a low computational complexity that can be used for networks where the topology continuously changes, and thus,  users' clustering and UAVs' positioning need to be regularly updated.
Finally, by providing various 3D simulations, we evaluate the effect of various system parameters such as the number and heights of UAVs, as well as the density and height of 3D obstacles on the LoS coverage.
\end{abstract}
\IEEEpeerreviewmaketitle

\section{Introduction}
\IEEEPARstart{W}{ith} the massive deployment of small cell base stations (SBSs) which are vital for the ongoing network densification, flexible, reliable, and easy-to-deploy wireless fronthaul links are required. 
Due to the suffering from low data rates of microwave bands, high-frequency millimeter wave (mmWave), terahertz (THz), and free-space optical (FSO) links are nominated as potential solutions for wireless fronthaul links to offload the explosive data traffic generated by massive deployments of SBSs.
However, mmWave/THz/FSO links require a Line of sight (LoS)  connection, which is the main hurdle in urban regions due to the existence of high and dense buildings.
A scalable idea is to adopt unmanned aerial vehicles (UAVs) as wireless fronthaul hub points to provide LoS connectivity which enable the implementation of mmWave/THz/FSO links in commercial systems \cite{dabiri2022modulating}.
In other words, due to their maneuverability and flexibility, UAVs should be positioned above tall buildings to provide LoS coverage for the SBSs/nodes distributed among the 3D obstacles, which is the main goal of this work.

\subsection{Literature Review}
Due to the high flying altitudes and maneuverability, UAVs are more likely to create LoS links, and therefore have the potential to establish higher capacity networks compared to terrestrial links.
Although numerous great works have been reported in the literature about UAV-based networks  \cite{8247211,8918497,9457160,9800925}, most of these works are based on LoS probability analysis and cannot be directly used for high-frequency directional THz/FSO antennas \cite{gemmi2022cost,8478112,gemmi2022properties}. Unlike lower frequency, directional high-frequency THz/FSO links do not have the ability to pass through obstacles where transmitted signals are severely weakened or interrupted. 

Therefore, analysis based on LoS probability does not lead to accurate results here. In other words, for high-frequency THz/FSO communications, it must be precisely determined whether the link is in LoS state or not.
More recently, special attention has been paid to the analysis of UAV-based networks under the presence of 3D obstacles in \cite{hu2020low, 9709500,
	zeng2021toward,zeng2021simultaneous,
	9200666,
	wang2020placement,sabzehali20213d,
	9044827, 
	lin2021adaptive,li2022geometric,
	%
	yi2022joint,yi20233,
	tang2021performance,zhu2022geometry}.
In \cite{hu2020low}, using the ray-tracing model, a UAV-aided relay network is studied where a UAV could fly to provide coverage for a number of ground users in an urban area with many obstructions. 
In \cite{9709500}, a novel binary channel concept is proposed to specify the LoS state of users and then use the binary channel for energy delivery.
The concept of channel knowledge map as an enabler toward environment-aware wireless communications is described in \cite{zeng2021toward}. 
A novel coverage-aware navigation approach is proposed in \cite{zeng2021simultaneous} to achieve ubiquitous 3D communication coverage for the UAVs in the sky.
In \cite{9200666}, authors try to create a LoS path between the source and the destination by adjusting the location of the UAV in relation to the 3D obstacles. However, the results of this work are limited to providing the LoS coverage for only two nodes.
By establishing the 3D directional coverage model, 3D placement of UAVs to provide LoS coverage for users has been studied in \cite{wang2020placement,sabzehali20213d}.
For simplicity, in \cite{wang2020placement} and \cite{sabzehali20213d}, a spherical base cone model is used for UAV LoS coverage, while in real scenarios, the LoS coverage is a function of the shape of the 3D obstacles which are mainly tall buildings. 
Also, even if the shape of the 3D obstacles is fixed, when the UAV's position changes, the 3D function of the LoS coverage changes, and thus, assigning a fixed model for the LoS coverage, although makes analysis and optimization easier, can be considered inaccurate.
%


As a different approach, a novel geometrical-based LoS model is proposed in \cite{9044827} to detect the blockage in the UAV-based communication system. 
Inspired by \cite{9044827}, in the last two years, valuable works have been proposed in \cite{lin2021adaptive,li2022geometric,
	yi2022joint,yi20233,
	tang2021performance,zhu2022geometry}.
An information collection method is proposed in \cite{lin2021adaptive}  based on the communication probability of LoS to solve the coverage problem of a UAV-aided network.
A new UAV deployment method is examined in \cite{li2022geometric} based on building geometric analysis. 
In \cite{yi2022joint} and \cite{yi20233}, the authors modeled the blockage effect caused by buildings according to the 3D geographical information to deploy multiple UAV  base stations (BSs).
An interference management algorithm is proposed in \cite{tang2021performance} via power control under 3D blockage effects. 
Using the statistical geographic information, a stochastic LoS probability model is proposed in \cite{zhu2022geometry} for 3D UAV-to-ground channels. 
In these works, by using the geometric characteristics of obstacles, an attempt is to determine the LoS state of the distributed users. The results of these works are suitable for lower frequencies or users with omnidirectional antennas, where a non-line-of-sight (NLOS) user might be connected by adapting transmit power. However, for a set of applications that are based on FSO or directional THz links, creating a connection is impossible in the case of NLoS due to the high attenuation.
%

In some applications of UAV-based THz/FSO links, optimization should be conducted in such a way to ensure that all distributed nodes among the 3D obstacles are covered by the UAVs. On the other hand, for a time-varying dynamic network, UAVs must be ready to provide communication links anywhere in the target area at any time. Thus, a set of UAVs should create LoS coverage for the entire target area or at least for a high percentage (nearly 100 percent). Therefore, in order to optimize the considered THz/FSO network, an algorithm must be provided to determine the LoS status of all points on the target area. In this case, the computational load is much higher than determining the LoS coverage for only several users, and thus, for optimizing such a network, a fast and reliable algorithm is necessary, which is the main goal of this work.




\subsection{Contribution}
In this work, inspired by \cite{lin2021adaptive,li2022geometric,
	yi2022joint,yi20233,
	tang2021performance,zhu2022geometry}, our target is to locate a set of UAVs equipped with directional THz links to provide LoS connectivity for a higher percentage of the target area.
To this end, we divide this work into two parts. In the first part, we specifically focus on the LoS coverage modeling. Then, using the 3D LoS model, we analyze a THz UAV-based communication network with 3D obstacles. In summary, the key contributions of this paper are summarized as follows:
\begin{itemize}
	\item First, we characterize and model the environment with 3D blocks. Then, we propose a user-friendly algorithm based on 3D spatial vectors which is able to determine the LoS state for all points within the target area. 
	\item Second, using 3D modeling, several positioning algorithms are provided to locate the set of UAVs in the 3D space in order to maximize the LoS coverage percentage in the target area. The performance of the positioning algorithms are compared in terms of the induced computational complexity.
	\item In the second part of our work, we focus on the optimal design of a UAV-based THz communication network to create LoS coverage for all randomly distributed nodes among 3D obstacles by optimally positioning the set of UAVs. To this end, a geometrical analysis-based algorithm is proposed, which optimally clusters the distributed nodes and locates the set of UAVs to optimize the average network capacity while ensuring that all the nodes are in the LoS state.
	\item A sub-optimal hybrid algorithm based on k-means and geometrical analysis of 3D obstacles is proposed for nodes clustering and UAVs positioning. The results of the proposed sub-optimal algorithm converges to the optimal values in a fast rate. The proposed sub-optimal algorithm can be used for networks where the network topology continuously changes, which requires a regular update in the nodes' clusters and the UAVs' positions.
	\item Finally, by providing various 3D simulations, we examine the performance of different optimization algorithms. We also evaluate the effect of various system parameters such as the number and heights of UAVs, as well as the density and height of 3D obstacles on the LoS coverage and the average capacity of the network.
\end{itemize}


\begin{figure}
	\begin{center}
		\includegraphics[width=3.4 in]{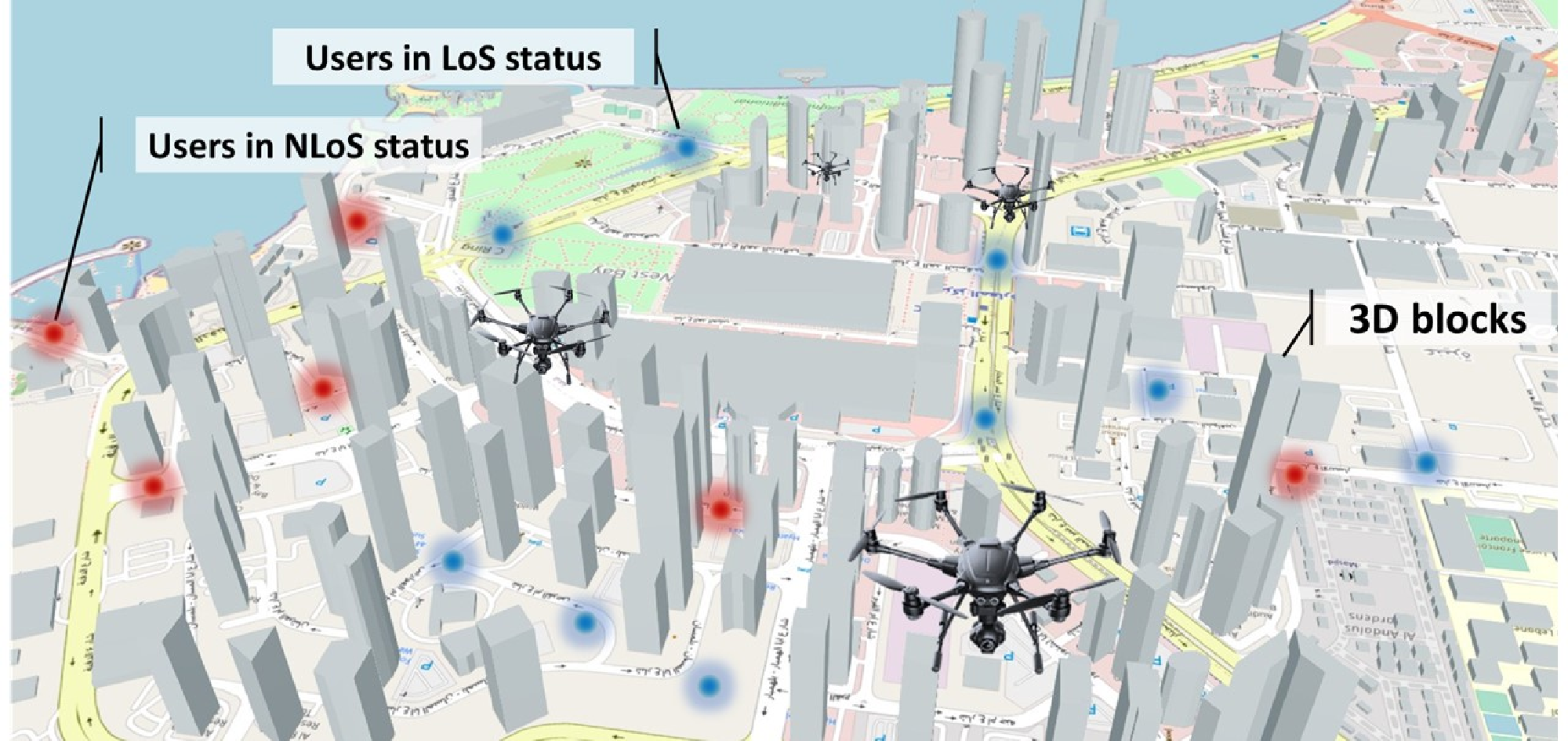}
		\caption{Graphical example of Doha, Qatar, West Bay towers where the dominant 3D obstacles are cube-shaped buildings \cite{dOHA_WE}.}
		\label{sm1}
	\end{center}
\end{figure}
%

\section{System Model and Main Assumptions}
In this section, we present our definition of the LoS coverage, followed by the framework to model an environment with 3D obstacles.


\subsection{LoS Coverage Definition}
\begin{figure}
	\begin{center}
		\includegraphics[width=3.4 in]{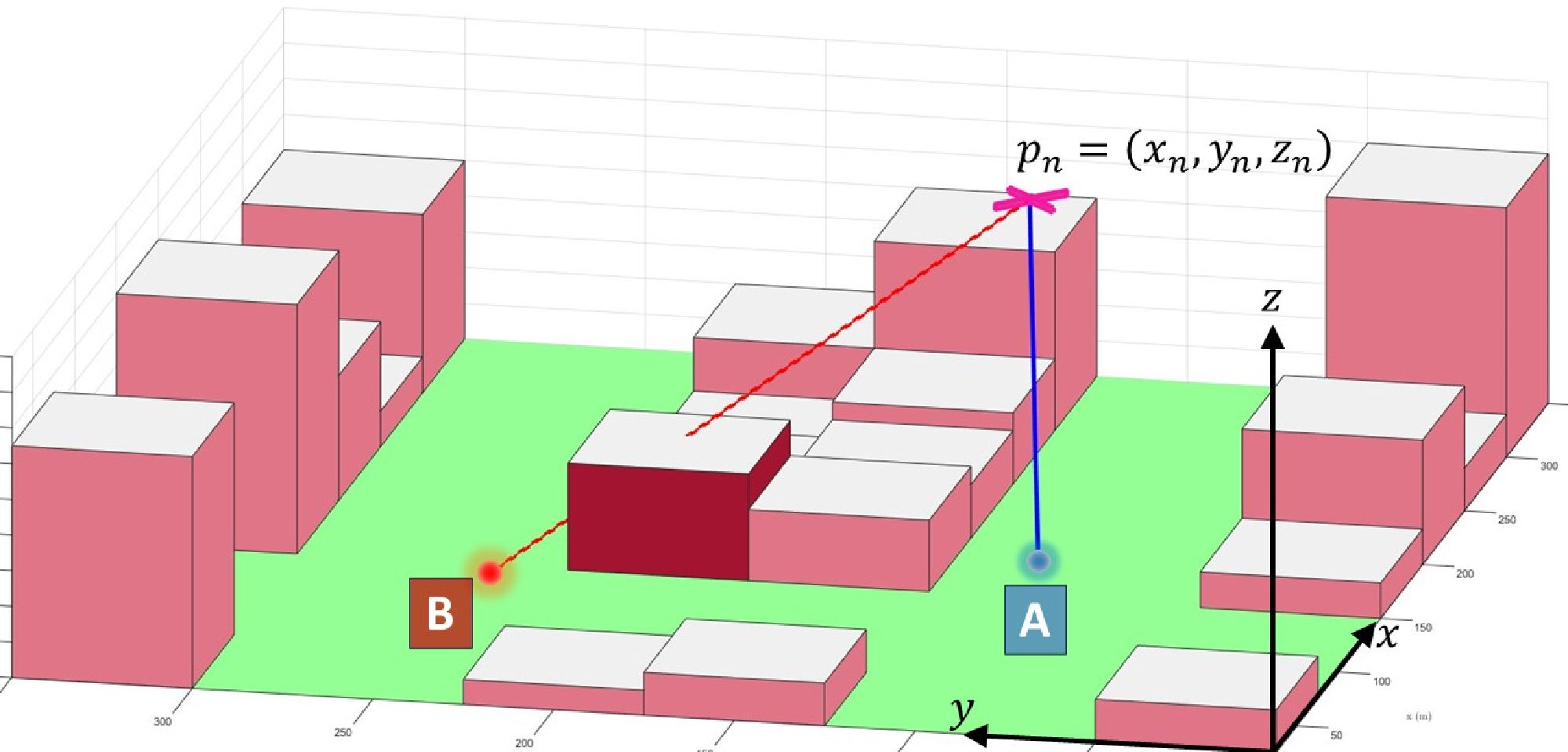}
		\caption{A graphical example of LoS and NLoS links. 
		Node A is in LoS state and node B is in NLoS state and they always stay in these states (with probability one) unless the UAV's position changes. }
		\label{sg1}
	\end{center}
\end{figure}

%
The analysis based on the LoS probability cannot be a suitable metric for THz and optical communications. To get a better view, a graphical example is provided in Fig. \ref{sg1} where we have two nodes A and B. Node A is in LoS state and node B is in NLoS state and they always stay in these states (with probability one) unless the topology changes, for example, if the position of the UAV or ground nodes changes.
In this work, as an approach towards more realistic results, instead of using the LoS probability, we use LoS coverage which is a more practical performance metric for THz/FSO communications. 

In some applications, the users/nodes' positions are not known, and a user may request access to the UAVs at any time and anywhere. Therefore, we need to put all the target area or at least a high percentage of the area in the LoS state by properly positioning the UAVs. In this case, we have a general definition of the LoS coverage as follows.

{\bf Definition 1.} {\it LoS coverage is the percentage of the area with LoS status}.

On the other hand, for some applications, the following definition of LoS coverage can be used.

{\bf Definition 2.} {\it Under the presence of 3D obstacles, the LoS coverage is defined as the percentage of ground users/nodes that are placed in the LoS state of the UAV.} 




\subsection{General System Parameters}
For the considered system, the UAVs in service are denoted by $u_n$ for $n\in\{1,...,N_u\}$, where $N_u$ is the number of UAVs.
The position of $u_n$ is located at $p_n=[x_n,y_n,z_n]$ in Cartesian coordinate system
$[x,y,z]\in\mathbb{R}^{1\times3}$, where $z_n<h_\text{max}$ is the height of the $n$th UAV, and $h_\text{max}$ is the maximum allowed height of the UAVs.
We also denote the target area to be covered by the UAVs with $S=d_x\times d_y$, which is a rectangular area with sides $[d_x,d_y]$ in the $x-y$ plane.
Moreover, let $B_m$ represents the $m$th 3D block for $m\in\{1,...,M_b\}$, where $M_b$ is the number of the 3D blocks.
Each block has its own Cartesian coordinate system denoted by $x'_m,y'_m,z'_m$, which is a coordinate plane rotated by $\theta_m$ around the $z$ axis. Therefore, the $z'_m$ axis of all the blocks is in the same direction ($z'_m=z$ for all $m\in\{1,...,M_b\}$) and only their $[x'_m,y'_m]$  axes are different.

Let $[x_{bm}, y_{bm}]$ represent the center of the $m$th obstacle in the $x-y$ coordinate plane. 
Considering that the dominant shape of buildings is a rectangular cube, in this work, we first model each 3D block as a rectangular cube with height $h_{bm}$ and sides $(d_{xm},d_{ym})$ in $[x'_m,y'_m]$ coordinate plane. 
The specifications of each a rectangular cube block include seven parameters as:
\begin{align}
	\label{s1}
	B_m = \left( \theta_m, x_{bm}, y_{bm}, z_{bm}, d_{xm}, d_{ym}, h_{bm}  \right).
\end{align}
where  $z_{bm}$ is the height of the ground level at the location of the $m$th block.

\begin{table}
	\textcolor{black}{
		\caption{The list of main notations.} 
		\centering 
		\begin{tabular}{l l} 
			\hline\hline \\[-1.2ex]
			{\bf Parameter} & {\bf Description}  \\ [.5ex] 
			\hline\hline \\[-1.2ex]
			\multicolumn{2}{l}{ \textbf{Target area to provide LoS coverage} } \\[-1ex]
			\multicolumn{2}{l}{ ---------------------------------------------------------------------------- } \\
			$[x,y,z]$ & $\in\mathbb{R}^{1\times3}$ is the main Cartesian coordinate system \\
			$S$ & $=d_x\times d_y$ is the target area \\
			$N_x\times N_y$ & Number of grid cells in target area \\
			$s_{ij}$ & $=d_{xi}\times d_{yj}$ denote each grid cell \\
		    $d_{xi}$ & $=d_x/N_x$ \\
		    $d_{yj}$ & $=d_y/N_y$ \\
		    $p_{ij}$ & $=[x_{ij},y_{ij},h_{ij}]$ where $x_{ij} = i d_{xi}$, $y_{ij}=j d_{yj}$ \\
		    $h_{ij}$ & Height of the ground level at $p_{ij}$ relative to the $s_{11}$ \\
		    $s_{11}$ & Considered as reference point in $S$ with $h_{11}=0$\\
			\multicolumn{2}{l}{ ---------------------------------------------------------------------------- } \\
			\multicolumn{2}{l}{ \textbf{3D obstacles:} From \eqref{s1}, it is characterized by 7 parameters } \\[-1ex]
			\multicolumn{2}{l}{ ---------------------------------------------------------------------------- } \\
			$M_b$ & Number of the blocks in the target area \\
			$B_m$ & Represents the $m$th 3D block \\ 
			$[x'_m,y'_m,z'_m]$ & Cartesian coordinate system of $B_m$, see Fig. \ref{ch1} \\
			$\theta_m$ & Rotation angle of $[x'_m,y'_m]$, see Fig. \ref{ch1} \\
			$x_{bm}, y_{bm}$ & Center of $B_m$ in the $[x,y]$ coordinate plane \\
			$z_{bm}$ & Height of the ground level at $B_m$\\
			$p_{bm}$ & $=[x_{bm},y_{bm},z_{bm}]$ \\
			$d_{xm}, d_{ym}$ & Side lengths of $B_m$\\
			$h_{bm}$  & Height of $B_m$\\
			%
			\multicolumn{2}{l}{ --------} \\
			\multicolumn{2}{l}{ \textbf{UAVs} } \\[-1ex]
			\multicolumn{2}{l}{ --------} \\
			$N_u$ & Number of UAVs \\
			$u_n$ & Denotes $n$th UAV\\
			$p_n$ & $=[x_n,y_n,z_n]$, position of $u_n$\\
			${\bf p}_u$ & $=[p_1,...,p_{N_u}]$ \\
			$h_\text{max}$ & Maximum allowed height of the UAVs, $z_n\leq h_\text{max}$ \\
			$A$ & $=d_x\times d_y$ is the area for UAVs  parallel to $S$\\
			$N_{ux}\times N_{uy}$ & Number of grid cells in $A$ \\
			$a_{ij}$  & $=d_{uxi}\times d_{uyj}$ denote each grid cell in $A$\\
			$d_{uxi}$ & $=d_x/N_{ux}$ \\
			$d_{uyj}$ & $=d_y/N_{uy}$ \\
			$p_{aij}$ & $=[x_{ij},y_{ij},h_u]$ where $x_{ij} = i d_{uxi}$, $y_{ij}=j d_{uyj}$ \\
	        \multicolumn{2}{l}{ ------------------------------------------------------------------------------------------ } \\
			\multicolumn{2}{l}{ \textbf{THz-based Network:} Additional parameters related to the definition 2 } \\[-1ex]
			\multicolumn{2}{l}{ ------------------------------------------------------------------------------------------ } \\
			$g_k$ & Represent the $k$th ground node where $k\in\{1,...,N_g\}$ \\
			$N_g$ & Represent the number of distributed ground nodes \\
			$p_{gk}$ & $=[x_k,y_k,h_k]$ indicates the location of the $g_k$\\
			${\bf p}_{g}$ & $=[p_{g1}, ...,p_{gN_g}]$ \\
			$h_{nk}$ & THz channel between $u_n$ and $g_k$ \\
			$h_{l,nk}$ & Path loss between $u_n$ and $g_k$ \\
			$G_k$ &  Array antenna gain of $g_k$ \\
			$G_n$ & Array antenna gain of $u_n$ \\
			$L_{nk}$ & Link length between $u_n$ and $g_k$ \\
			\hline \hline              
	\end{tabular} }
	\label{I1} 
\end{table} 

For coverage analysis, we first partition the target area $S=d_x\times d_y$ into $N_x\times N_y$ grid cells. Each grid cell is denoted by $s_{ij}=d_{xi}\times d_{yj}$ where $d_{xi}=d_x/N_x$, and $d_{yj}=d_y/N_y$.
The larger $N_x$ and $N_y$ are chosen, the grids $s_{ij}$ tend to a point in the $S$ plane, and the accuracy of the coverage analysis increases by paying more processing computation. The appropriate selection of $N_x$ and $N_y$ depends on the dimensions of $(d_x,d_y)$, the desired accuracy, and the maximum processing load that can be tolerated by the processor. Moreover, the position of each grid $s_{ij}$ is determined as $p_{ij}=[x_{ij},y_{ij},h_{ij}]$ in $[x,y,z]$ coordinate system, where $x_{ij} = i d_{xi}$, $y_{ij}=j d_{yj}$, and $h_{ij}$ is height of the ground level relative to the height of the grid $s_{11}$ which is considered as a reference point.

Now, we define the $N_x\times N_y$ coverage matrix ${\bf C}_n$ for the $n$th UAV where $c_{nij}$ represents the ($i,j$)th entry of ${\bf C}_n$.
%
$c_{nij}=1$ if the grid $s_{ij}$ is in the LoS of $u_n$ and $c_{nij}=0$ if the grid $s_{ij}$ is in the NLoS state.
We also define the matrix ${\bf C}'_n=1 - {\bf C}_n$, which represents the grids that are in NLoS state and obtained as:
\begin{align}
	\label{s3}
	{\bf C}'_n = \sum_{m=1}^{M_b} {\bf C}'_{nm},
\end{align}
where the matrix $ {\bf C}'_{nm}$ shows the areas of the coverage area $S$ that are placed in the NLoS state of the $u_n$ due to the $m$th block. $c'_{nmij}$ represents the ($i,j$)th entry of ${\bf C}'_{nm}$ where $c'_{nmij}=1$ if it is in the NLoS state due to the $m$th block and $c'_{nmij}= 0$ otherwise.
Moreover, $s'_{nij}$ represents the ($i,j$)th entry of ${\bf C}'_n$.

{\bf Remark 1.} {\it Some NLoS grids in $S$ may be caused by two or more blocks. In this case, we have $c'_{nij}>1$. For simplicity and without loss of generality, since the NLoS state caused by one or more blocks is not different, we find the entries with $c'_{nij}>1$ and set them as $c'_{nij}=1$.}

{\bf Remark 2.} {\it In this work, grid $s_{11}$ is considered as a reference point for height. Therefore, all the heights including the height of UAVs, blocks, and grids $s_{ij}$ are obtained relative to the $s_{11}$ reference point.}

\subsection{System Parameters Related to Definition 2}
According to definition 2, we have a distribution of ground nodes among the 3D blocks. Let $g_k$ represent the $k$th node where $k\in\{1,...,N_g\}$ and $N_g$ represent the number of nodes. Also, $p_{gk}=[x_k,y_k,h_k]$ indicates the location of the $g_k$.
Each UAV is equipped with $N_a$ directional THz antennas, and each directional antenna is a square array antenna with $N_e\times N_e$ antenna elements. 
It should be noted that the dimensions of the directional THz antennas are small and can be installed on the UAV. For a better view, we present a comparative example regarding the dimensions of the antenna at frequencies 2 GHz and 188 GHz. Based on \cite{balanis2016antenna,3gppf}, the dimensions of a simple square array antenna at frequency 2 GHz with $N_e\times N_e=20\times20$ antenna elements and $\lambda_c/2$ spacing between antenna elements is obtained theoretically as
$\left(20\times\frac{\lambda_c}{2}\right)^2
=\left(20\frac{c=3\times10^8}{f_c=2\times10^9}\right)^2
=(1.5)^2 ~\text{m}^2 $
where $\lambda_c$ is the wavelength. It is clear that the use of array antennas in low frequencies of around 1-10 GHz will face many challenges in practice due to their high dimensions and weight.
However, for the popular THz frequency 183 GHz, the dimensions of the simple square array antenna with $N_e\times N_e=20\times20$ antenna elements are reduced theoretically as
$\left(20\times\frac{\lambda_c}{2}\right)^2
=\left(20\frac{c=3\times10^8}{f_c=183\times10^9}\right)^2
\simeq (1.6)^2~\text{cm}^2$.
Therefore, due to the small dimensions, it is possible to use several THz array antennas on the UAV \cite{9998554}. 
The THz channel between $u_n$ and $g_k$ is modeled as \cite{boulogeorgos2019analytical}:
\begin{align}
	h_{nk} = h_{l,nk} \sqrt{G'_{kn} G_{nk}},
\end{align}
where $h_{l,nk}$ models the path loss between $u_n$ and $g_k$, and $G'_{kn}$ and $G_{nk}$ are the array antenna gains between $g_k$ and $u_n$ of ground and UAV nodes, respectively.
The path loss coefficient can be expressed as
\begin{align}
	\label{los1}
	h_{l,nk} = h_{l_\text{free}} h_{l_\text{mol}},
\end{align}
where $h_{l_\text{free}}=\frac{c}{4 \pi f_t L_{nk}}$ is the well-known free-space path loss, $L_{nk}$ is the link length between $u_n$ and $g_k$, and $f_t$ is the THz frequency.  
Also, $h_{l_\text{mol}}$ is the molecular absorption loss which is formally described by the Beer–Lambert law as \cite{kokkoniemi2021line}:
\begin{align}
	\label{los2}
	h_{l_\text{mol}} = \exp\left(- \frac{\left[\sum_i y_i(f_t,\mu) + g(f_t,\mu)  \right] L_{nk}}{2}   \right),
\end{align}
where $\mu=\frac{r_h}{100} \frac{p_\text{pres}(T,p)}{p}$ is the volume mixing ratio of water vapor, $r_h$ is the relative humidity, $T$ is the environment temperature in degrees centigrade, $p$ is the pressure in hectopascals, and $p_\text{pres}(T,p)$ is the saturated water vapor partial pressure formulated in \cite{alduchov1996improved}.
Also, $y_i(f_t,\mu)$ formulates six polynomials for six major absorption lines centered at 119, 183, 325, 380, 439, and 448 GHz frequencies as \cite{kokkoniemi2021line}:
\begin{align}
	\label{los3}
	y_i(f_t,\mu) = \frac{A_i(\mu)}{B_i(\mu) + (\frac{f_t}{100c}-p_i)^2}, 
\end{align}
where
\begin{align}
	\left\{ \!\!\!\!\! \! \!
	\begin{array}{rl}
		& A_1(\mu) = 5.159\times 10^{-5}(1 - \mu)(-6.65 \\
		&~~~~~~~~~~\times 10^{-5}(1 - \mu) + 0.0159), \\
		& A_2(\mu) = 0.1925\mu (0.1350\mu + 0.0318),\\
		& A_3(\mu) = 0.2251\mu(0.1314\mu + 0.0297),\\
		& A_4(\mu) = 2.053\mu(0.1717\mu + 0.0306),\\
		& A_5(\mu) = 0.177\mu(0.0832\mu + 0.0213),\\
		& A_6(\mu) = 2.146\mu(0.1206\mu + 0.0277),
	\end{array} \right. \nonumber
\end{align}
and
\begin{align}
	\left\{ \!\!\!\!\! \! \!
	\begin{array}{rl}
		& B_1(\mu) = (-2.09\times 10^{-4}(1 -\mu) + 0.05)^2, \\
		& B_2(\mu) = (0.4241\mu + 0.0998)^2, \\
		& B_3(\mu) = (0.4127\mu + 0.0932)^2, \\
		& B_4(\mu) = (0.5394\mu + 0.0961)^2, \\
		& B_5(\mu) = (0.2615\mu + 0.0668)^2, \\
		& B_6(\mu) = (0.3789\mu + 0.0871)^2. \\
	\end{array} \right. \nonumber
\end{align}
Also $p_1 = 3.96$ cm$^{-1}$, $p_2=6.11$ cm$^{-1}$, $p_3=10.84$ cm$^{-1}$, $p_4=12.68$ cm$^{-1}$, $p_5=14.65$ cm$^{-1}$, $p_6=14.94$ cm$^{-1}$.
Moreover, $g(f_t,\mu)$ is a polynomial to fit the expression to the actual theoretical response;
\begin{align}
	g(f_t,\mu) = \frac{\mu}{0.0157} \left( 2\times 10^{-4} +0.915\times10^{-112}f_t^{9.42}\right).
\end{align}


\begin{figure*}
	\centering
	\subfloat[] {\includegraphics[width=3.8 in]{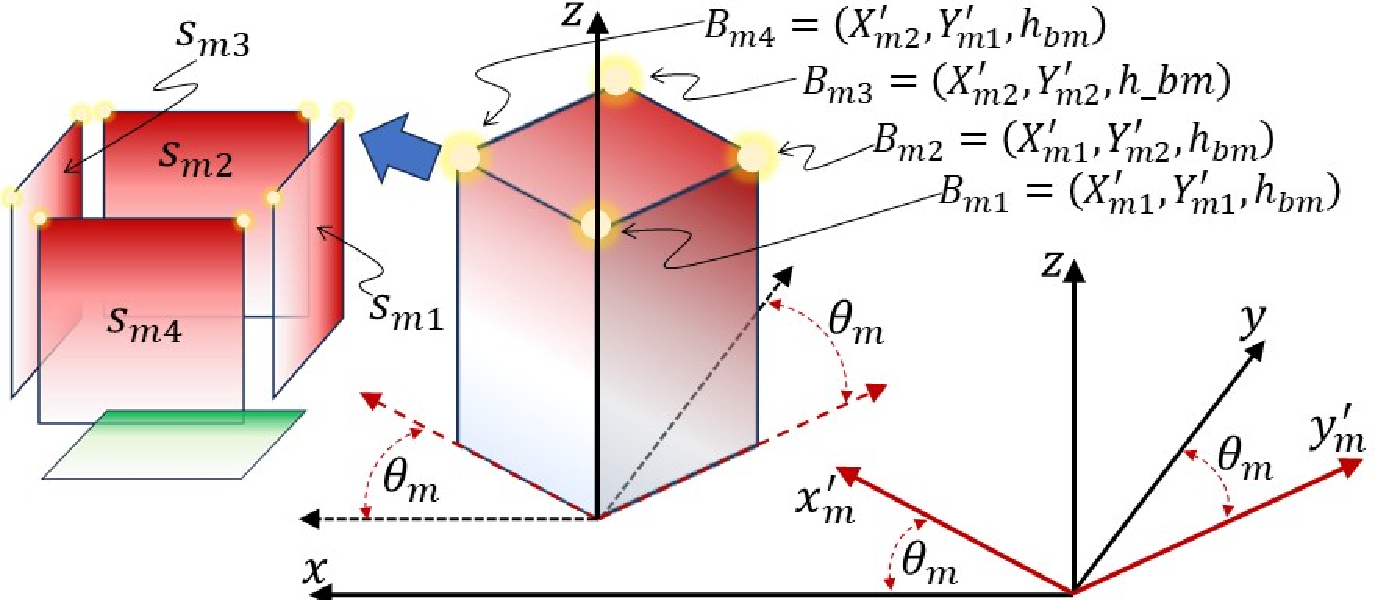}
		\label{ch1}
	}
	\hfill
	\subfloat[] {\includegraphics[width=3.2 in]{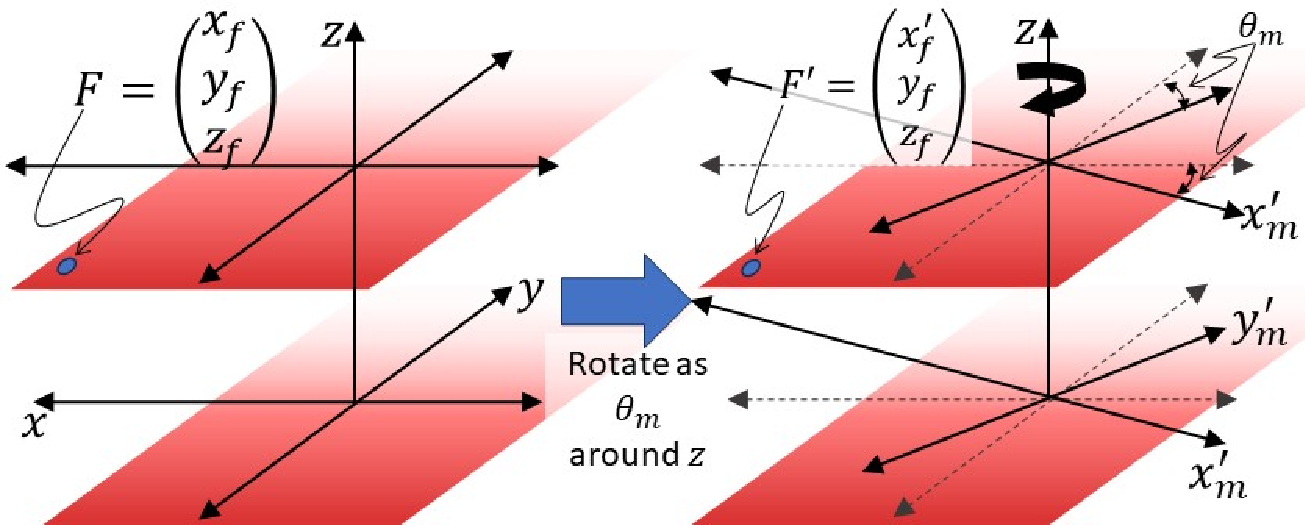}
		\label{ch2}
	}
	\caption{(a) Illustrating the main $x-y$ coordinate plane of the target area $S$ and the $x'_m-y'_m$ coordinate plane of the $m$th block along with specifying the key characteristics of each block used in Algorithm 1.
		(b) Illustrating that the $x'_m-y'_m$ coordinate plane is obtained by rotating the main $x-y$ coordinate plane around the $z$ axis by $\theta_m$.
	}
	\label{ch}
\end{figure*}
%

\section{LoS Coverage Modeling under 3D obstacles}
Due to the high altitude of UAVs, short-height obstacles can be ignored with good accuracy, and thus, the dominant obstacles in cities are tall buildings. 
Considering that the dominant shape of buildings is a rectangular cube, in this work, we first model the LoS coverage for rectangular cube blocks. 
Then, we extend the LoS coverage model for several other possible block shapes. 

\subsection{LoS Coverage Model for Cube Blocks}
In this section, we model the LoS coverage based on definition 1. Although the system model related to definition 1 seems simpler, the dimensions of matrices are large, which requires more computational load during the optimization and positioning of UAVs. Therefore, it is necessary to provide a fast algorithm to optimize and calculate coverage percentage.
To this end, we first model the LoS coverage for the $m$th block, which can be repeated for the rest of the blocks, in a similar way.
To model the matrix ${\bf C}_{nm}$ resulting from the $m$th block and $n$th UAV, we perform the following steps.
\subsubsection{Step 1}
We convert the coordinates of all the points in the $x-y$ coordinate plane to the $x'_m-y'_m$ coordinate plane of the $m$th block. The $x'_m-y'_m$ coordinate system is the result of rotating the $x-y$ coordinate system equal to $\theta_m$ around the $z$ axis. As shown in Fig. \ref{ch}, there is a one-to-one relation between point ${\mathbf{f}}=[x_f,y_f,z_f]$ in the $x-y$ coordinate system and ${\mathbf{f}}'$ in the $x'_m-y'_m$ coordinate system as:
\begin{align}
	\label{s4}
	\underbrace{\left( \begin{matrix} 
		x_f  \\
		y_f   \\
		z_f  \\
	\end{matrix}  \right)}_{{\mathbf{f}}}
    = 
    \underbrace{ \left( \begin{matrix} 
    	\cos(\theta_m) & -\sin(\theta_m) & 0  \\
    	\sin(\theta_m) & \cos(\theta_m) & 0  \\
    	0 & 0 & 1  \\
    \end{matrix}  \right) }_{{\mathbf{R}}} 
    \underbrace{\left( \begin{matrix} 
    	x'_f  \\
    	y'_f   \\
    	z'_f   \\
    \end{matrix}    \right)}_{{\mathbf{f}}'}.
\end{align}
We can also represent \eqref{s4} as
\begin{align}
	\label{s5}
	{\mathbf{f}}' = {\mathbf{R}}^{-1} {\mathbf{f}},
\end{align}
where ${\mathbf{R}}^{-1}$ is the inverse matrix of ${\mathbf{R}}$. 
Using \eqref{s4}, the position of the $n$th UAV ${\mathbf{p}}_n=[x_n,y_n,z_n]$ for $n\in\{1,...,N_u\}$, the position of the $m$th block ${\mathbf{p}}_{bm}=[x_{bm},y_{bm},z_{bm}]$, and the positions of grids ${\mathbf{p}}_{ij}=[x_{ij},y_{ij},h_{ij}]$ for $i\in\{1,...,N_x\}$ and $j\in\{1,...,N_y\}$ are obtained in the $x'_m-y'_m$ coordinate system as:
\begin{align} \label{ss4}
	\left\{ \!\!\!\!\! \! \!
	\begin{array}{rl}
		& {\mathbf{p}}'_n = {\mathbf{R}}^{-1} {\mathbf{p}}_n, ~~\text{where}~~{\mathbf{p}}'_n=[x'_n,y'_n,z_n],\\
		& {\mathbf{p}}'_{bm} = {\mathbf{R}}^{-1} {\mathbf{p}}_{bm}, ~~\text{where}~~{\mathbf{p}}'_{bm}=[x'_{bm},y'_{bm},z_{bm}],\\
		& {\mathbf{p}}'_{ij} = {\mathbf{R}}^{-1} {\mathbf{p}}_{ij}, ~~\text{where}~~{\mathbf{p}}'_{ij}=[x'_{ij},y'_{ij},h_{ij}].\\
	\end{array} \right. 
\end{align}

\subsubsection{Step 2}
In this step, we obtain the set of 3D spatial vectors between the $n$th UAV and the set of grids $s_{ij}$. The 3D spatial vector is formulated as:
\begin{align}
	\label{s6}
	\frac{x'_m - x'_{n}}{v_{xij}} = 
	\frac{y'_m - y'_{n}}{v_{yij}} =
	\frac{z_m - z_{n}}{v_{zij}},
\end{align}
where
$v_{xij} = x'_{n}-x'_{ij}$, 
$v_{yij} = y'_{n}-y'_{ij}$, and
$v_{zij} = z_{n}-h_{ij}$. 
To increase the speed of calculations, we define $N_x\times N_y$ matrices ${\bf V}_{x}$, ${\bf V}_{y}$, and ${\bf V}_{z}$, whose entries are $v_{xij}$, $v_{yij}$, and $v_{zij}$, respectively. 
Also, by inverting the entries of the matrix ${\bf V}_{x}$,
we define $N_x\times N_y$ matrix ${\bf V}'_{x}$ as
\begin{equation} 
	{\bf V}'_{x} = \begin{pmatrix} 
		v_{x11}^{-1} & v_{x12}^{-1} & \cdots & v_{x1N_y}^{-1} \\
		v_{x21}^{-1} & v_{x22}^{-1} & \cdots & v_{x2N_y}^{-1} \\ 
		\vdots & \vdots & \vdots & \vdots\\ 
		v_{xN_x1}^{-1} & v_{xN_x2}^{-1} & \cdots & v_{xN_xN_y}^{-1} 
	\end{pmatrix}. 
\end{equation}
Similar to the ${\bf V}'_{x}$, we define the matrices ${\bf V}'_{y}$ and ${\bf V}'_{z}$ by inverting the entries of the matrices ${\bf V}_{y}$ and ${\bf V}_{z}$, respectively.

\begin{algorithm}
	\caption{LoS Coverage Modeling under 3D Obstacles}
	\begin{algorithmic}[1]
		\renewcommand{\algorithmicrequire}{\textbf{Input:}}
		\renewcommand{\algorithmicensure}{\textbf{Output:}}
		\REQUIRE $S=d_x\times d_y$, $N_x$, $N_y$, $N_u$, $M_b$, $p_n$, $h_\text{max}$, and information about 3D obstacles
		\ENSURE  ${\bf C}_n$ = 1 - ${\bf C}'_n$
	\FOR {m = 1 : $M_b$}
	\STATE  Convert the coordinates of $p_{bm}$ to $p'_{bm}$ using \eqref{s4}-\eqref{ss4}
	\STATE Convert the coordinates of $p_{ij}=[x_{ij},y_{ij},h_{ij}]$ to $p'_{ij}=[x'_{ij},y'_{ij},h_{ij}]$
	for all $i\in\{1,...,N_x\}$ and $j\in\{1,...,N_y\}$.
	\STATE Convert the coordinates of $p_n=[x_n,y_n,z_n]$ to $p'_n=[x'_n,y'_n,z_n]$. 
	for all $n\in\{1,...,N_u\}$
	\STATE Compute ${\bf V}_{x}$, ${\bf V}_{y}$, and ${\bf V}_{z}$ along with 
	${\bf V}'_{x}$, ${\bf V}'_{y}$, and ${\bf V}'_{z}$
	\STATE Compute ${\bf Y}_{X'_{m1}}$ and ${\bf Z}_{X'_{m1}}$ using \eqref{f1}
	\STATE Compute ${\bf X}_{Y'_{m2}}$ and ${\bf Z}_{Y'_{m2}}$ using \eqref{f1}
	\STATE Compute ${\bf Y}_{X'_{m2}}$ and ${\bf Z}_{X'_{m2}}$ using \eqref{f1}
	\STATE Compute ${\bf X}_{Y'_{m1}}$ and ${\bf Z}_{Y'_{m1}}$ using \eqref{f1}
	\FOR {$i=1:N_x$}
	\FOR {$j=1:N_y$}
	\IF {$Y'_{m1}<y_{xm1ij}<Y'_{m2}$ and $z_{bm}<z_{xm1ij}<h_{bm}$}
	\STATE Update $c'_{nij}=1$
	\ELSIF {$X'_{m1}<x_{ym2ij}<X'_{m2}$ and $z_{bm}<z_{ym2ij}<h_{bm}$}
	\STATE Update $c'_{nij}=1$
	\ELSIF {$Y'_{m1}<y_{xm2ij}<Y'_{m2}$ and $z_{bm}<z_{xm2ij}<h_{bm}$}
	\STATE Update $c'_{nij}=1$
	\ELSIF {$X'_{m1}<x_{ym1ij}<X'_{m2}$ and $z_{bm}<z_{ym1ij}<h_{bm}$}
	\STATE Update $c'_{nij}=1$
	\ENDIF
	%
	%
	\ENDFOR
	\ENDFOR
	\ENDFOR
\end{algorithmic} 
\end{algorithm}

\subsubsection{Step 3}
As shown in Fig. \ref{ch1}, the four upper vertices of the $m$th block are denoted as
\begin{align}
	\left\{ \!\!\!\!\! \! \!
	\begin{array}{rl}
		& B_{m1}=(X'_{m1},Y'_{m1},h_{bm}),~~
		X'_{m1}= x'_{bm} - d_{xm}/2, \\
		& B_{m2}=(X'_{m1},Y'_{m2},h_{bm}), ~~
		X'_{m2}= x'_{bm} + d_{xm}/2,\\
		& B_{m3}=(X'_{m2},Y'_{m2},h_{bm}), ~~
		Y'_{m1}= y'_{bm} - d_{ym}/2,\\
		& B_{m4}=(X'_{m2},Y'_{m1},h_{bm}), ~~
		Y'_{m2}= y'_{bm} + d_{ym}/2.\\
	\end{array} \right. 
\end{align} 
Since the blocks consist of 4 lateral faces (planes), in this step, we check the 4 lateral faces separately to see if the set of 3D spatial vectors obtained in step 2 collide with them or not. As shown in Fig. \ref{ch1}, the lateral face below the line connecting point $B_{m1}$ to $B_{m2}$ is called $s_{m1}$, and similarly, the lateral faces below lines connecting $B_{m2}$ to $B_{m3}$, $B_{m3}$ to $B_{m4}$, and $B_{m4}$ to $B_{m1}$ are called $s_{m2}$, $s_{m3}$, and $s_{m4}$ respectively. The main characteristic of all points in the $s_{m1}$ plane is that the $X'_{m1}$ is constant for all points. Using this point as well as \eqref{s6}, for $X'_{m1}$, we obtain the set of points on the 3D spatial vector in $x'_m-y'_m$ and $x'_m-z'_m$ planes as:
\begin{align}  
	\label{f1}
	\left\{ \!\!\!\!\! \! \!
	\begin{array}{rl}
	&{\bf Y}_{X'_{m1}} = {\bf V}'_{x}\odot{\bf V}_{y}(X'_{m1} - x'_n) + y'_n,  \\
	&{\bf Z}_{X'_{m1}} = {\bf V}'_{x}\odot{\bf V}_{z}(X'_{m1} - x'_n) + z_n,
    \end{array} \right.  %
\end{align}
where $\odot$ represents the Hadamard (element-wise) product.
Let $y_{xm1ij}$ and $z_{xm1ij}$ denote the ($i,j$)th entry of ${\bf Y}_{X'_{m1}}$ and ${\bf Z}_{X'_{m1}}$, respectively.
Also, $N_x\times N_y$ matrix $ {\bf C}'_{nm1}$ denotes the parts of the target area $S$ that are placed in the NLoS state of the $u_n$ due to the $s_{m1}$. 
$c'_{nm1ij}$ represents the ($i,j$)th entry of ${\bf C}'_{nm1}$ where $c'_{nm1ij}=1$ if it is in the NLoS state due to the $s_{m1}$ and $c'_{nm1ij}= 0$ otherwise.
Therefore, in order to obtain $ {\bf C}'_{nm1}$, the following conditions must be met for the points obtained in \eqref{f1}:
\begin{align}  
	\label{c1}
	&c'_{nm1ij}= \\
	&\left\{ \!\!\!\!\! \! \!
	\begin{array}{rl}
		& 1,~~~  
		Y'_{m1}<y_{xm1ij}<Y'_{m2}~\&~z_{bm}<z_{xm1ij}<h_{bm},\\
		& 0,~~~\text{otherwise}.
	\end{array} \right. \nonumber %
\end{align}
Similarly, we repeat these for the remaining three lateral faces as
\begin{align}  
	\label{f2}
	\left\{ \!\!\!\!\! \! \!
	\begin{array}{rl}
		&{\bf X}_{Y'_{m2}} = {\bf V}'_{y}\odot{\bf V}_{x}(Y'_{m2} - y'_n) + x'_n   \\
		&{\bf Z}_{Y'_{m2}} = {\bf V}'_{y}\odot{\bf V}_{z}(Y'_{m2} - y'_n) + z_n,
	\end{array} \right.
\end{align}
for $s_{m2}$ plane, 
\begin{align}  
	\label{f3}
	\left\{ \!\!\!\!\! \! \!
	\begin{array}{rl}
		&{\bf Y}_{X'_{m2}} = {\bf V}'_{x}\odot{\bf V}_{y}(X'_{m2} - x'_n) + y'_n   \\
		&{\bf Z}_{X'_{m2}} = {\bf V}'_{x}\odot{\bf V}_{z}(X'_{m2} - x'_n) + z_n,
	\end{array} \right.
\end{align}
for $s_{m3}$ plane, and
\begin{align}  
	\label{f4}
	\left\{ \!\!\!\!\! \! \!
	\begin{array}{rl}
		&{\bf X}_{Y'_{m1}} = {\bf V}'_{y}\odot{\bf V}_{x}(Y'_{m1} - y'_n) + x'_n   \\
		&{\bf Z}_{Y'_{m1}} = {\bf V}'_{y}\odot{\bf V}_{z}(Y'_{m1} - y'_n) + z_n.
	\end{array} \right.
\end{align}
for $s_{m4}$ plane, where
$x_{ym2ij}$, $y_{xm2ij}$, $x_{ym1ij}$, 
$z_{ym2ij}$, $z_{xm2ij}$, and $z_{ym1ij}$ are the ($i,j$)th entry of 
${\bf X}_{Y'_{m2}}$, ${\bf Y}_{X'_{m2}}$, ${\bf X}_{Y'_{m1}}$, 
${\bf Z}_{Y'_{m2}}$, ${\bf Z}_{X'_{m2}}$, and ${\bf Z}_{Y'_{m1}}$, respectively.
Similarly, to specify collisions with lateral planes $s_{m2}$, $s_{m3}$, and $s_{m4}$, the following conditions must be met, respectively:
\begin{align}  
	\label{c2}
	&c'_{nm2ij}= \\
	&\left\{ \!\!\!\!\! \! \!
	\begin{array}{rl}
		& 1,~~~  
		X'_{m1}<x_{ym2ij}<X'_{m2}~\&~z_{bm}<z_{ym2ij}<h_{bm},
		\\
		& 0,~~~\text{otherwise},
	\end{array} \right. \nonumber %
\end{align}
\begin{align}  
	\label{c3}
	&c'_{nm3ij}= \\
	&\left\{ \!\!\!\!\! \! \!
	\begin{array}{rl}
		& 1,~~~  
		Y'_{m1}<y_{xm2ij}<Y'_{m2}~\&~z_{bm}<z_{xm2ij}<h_{bm},\\
		& 0,~~~\text{otherwise},
	\end{array} \right. \nonumber %
\end{align}
\begin{align}  
	\label{c4}
	&c'_{nm4ij}= \\
	&\left\{ \!\!\!\!\! \! \!
	\begin{array}{rl}
		& 1,~~~  
		X'_{m1}<x_{ym1ij}<X'_{m2}~\&~z_{bm}<z_{ym1ij}<h_{bm},\\
		& 0,~~~\text{otherwise}.
	\end{array} \right. \nonumber %
\end{align}
Finally, using the proposed steps, we obtain the matrix ${\bf C}_n$ by Algorithm 1. According to the inputs including the information and specifications of the 3D blocks of an environment, Algorithm 1 quickly calculates the LoS matrix and determines which points of the target area $S$ are in the LoS state of the $u_n$ and which points are in the NLoS state. Due to the fast speed of Algorithm 1, in the following sections, it is used for optimization and positioning of UAVs to increase the LoS coverage percentage of the target area $S$.

%
\begin{figure*}
	\centering
	\subfloat[] {\includegraphics[width=3.4 in]{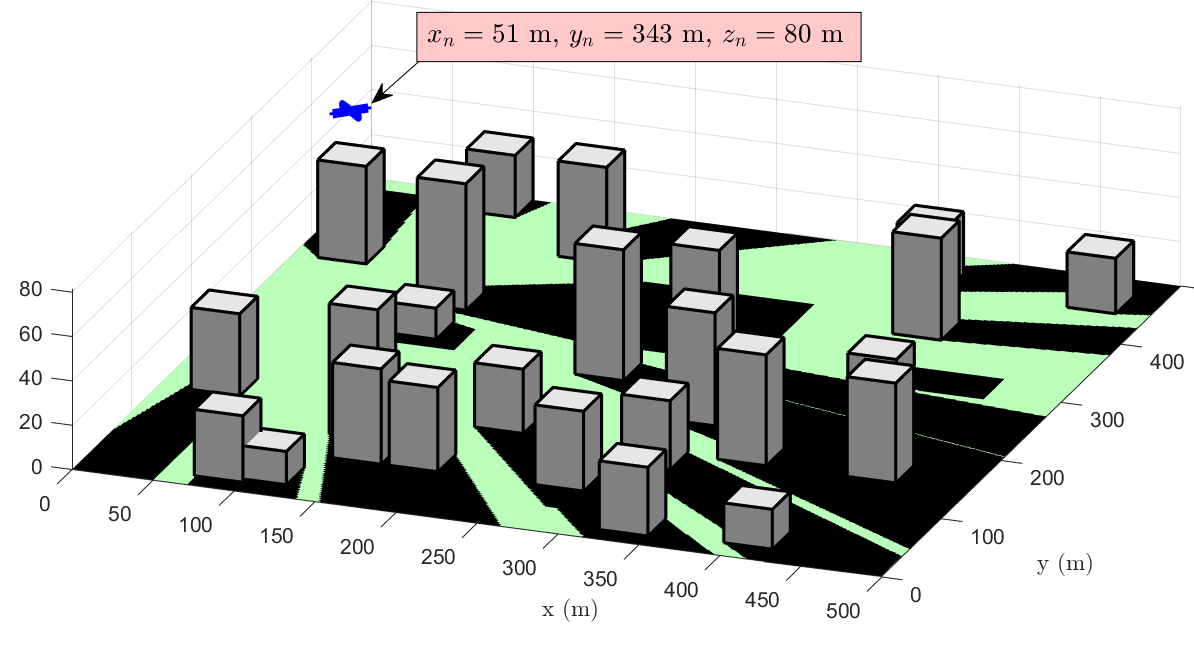}
		\label{cn1}
	}
	\hfill
	\subfloat[] {\includegraphics[width=3.4 in]{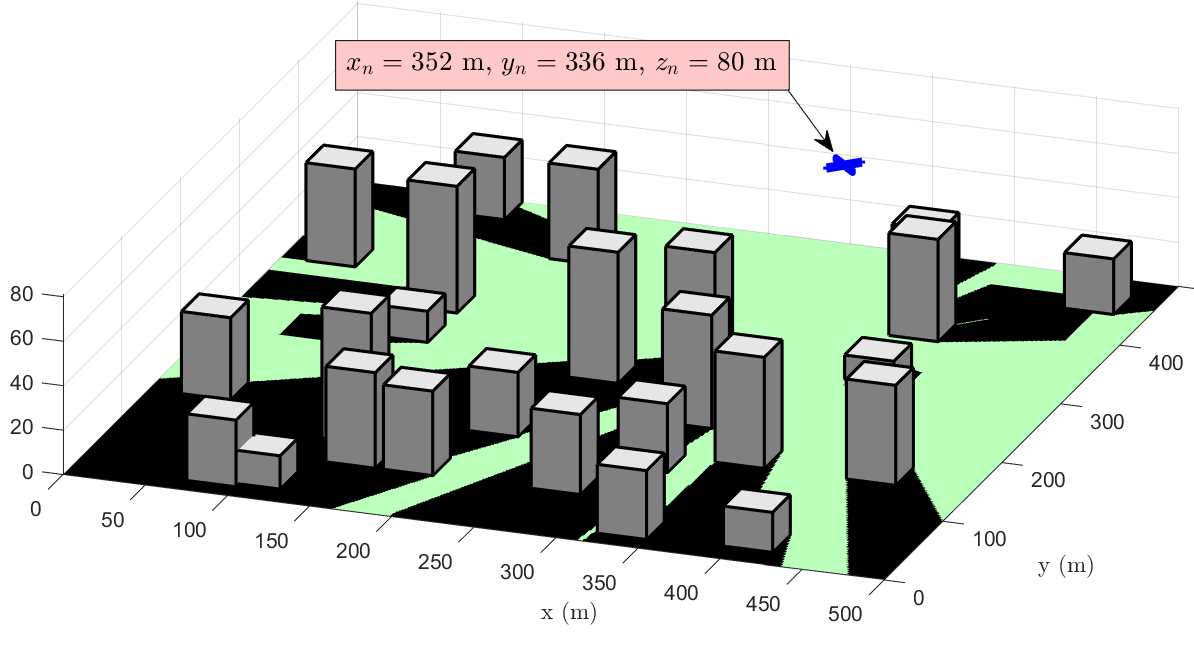}
		\label{cn2}
	}
	\caption{(a) An example of NLoS coverage obtained from Algorithm 1 for a random location $p_n=[51,343,80]$. (b) NLoS coverage changes by changing the position of the UAV to $p_n=[352,336,80]$. 
	}
	\label{cn}
\end{figure*}
%

For more readability, several loops are used in Algorithm 1. To increase the execution speed of Algorithm 1, it is recommended to use matrix operations instead of loops.
As an example, in Fig. \ref{cn1}, using Algorithm 1, we obtain the set of points in the target area that are in the LoS state. For simulation, a $500\times500$ m$^2$ environment is considered in which 25 buildings with random height and location are placed. The UAV is placed in the location $(x_n=53, y_n=343, z_n=80)$ m.
The shadows created represent the areas that are in the LoS state. The important point is that the LoS coverage is a function of the location of the UAV and as seen in Fig. \ref{cn2}, by changing the location of the UAV to $(x_n=352, y_n=336, z_n=80)$ m, the coverage of the target area changes, completely. Therefore, finding the optimal location for a set of UAVs is of great importance to reach the maximum LoS coverage. Moreover, the presented algorithm obtains ${\bf C}_n$ with high speed, regularly and purposefully. It is worth highlighting that the speed of calculations is very important for the optimal design of the system, especially when multiple UAVs are used to achieve a higher percentage of LoS coverage.

\begin{figure}
	\begin{center}
		\includegraphics[width=3.2 in, height= 2.2 in]{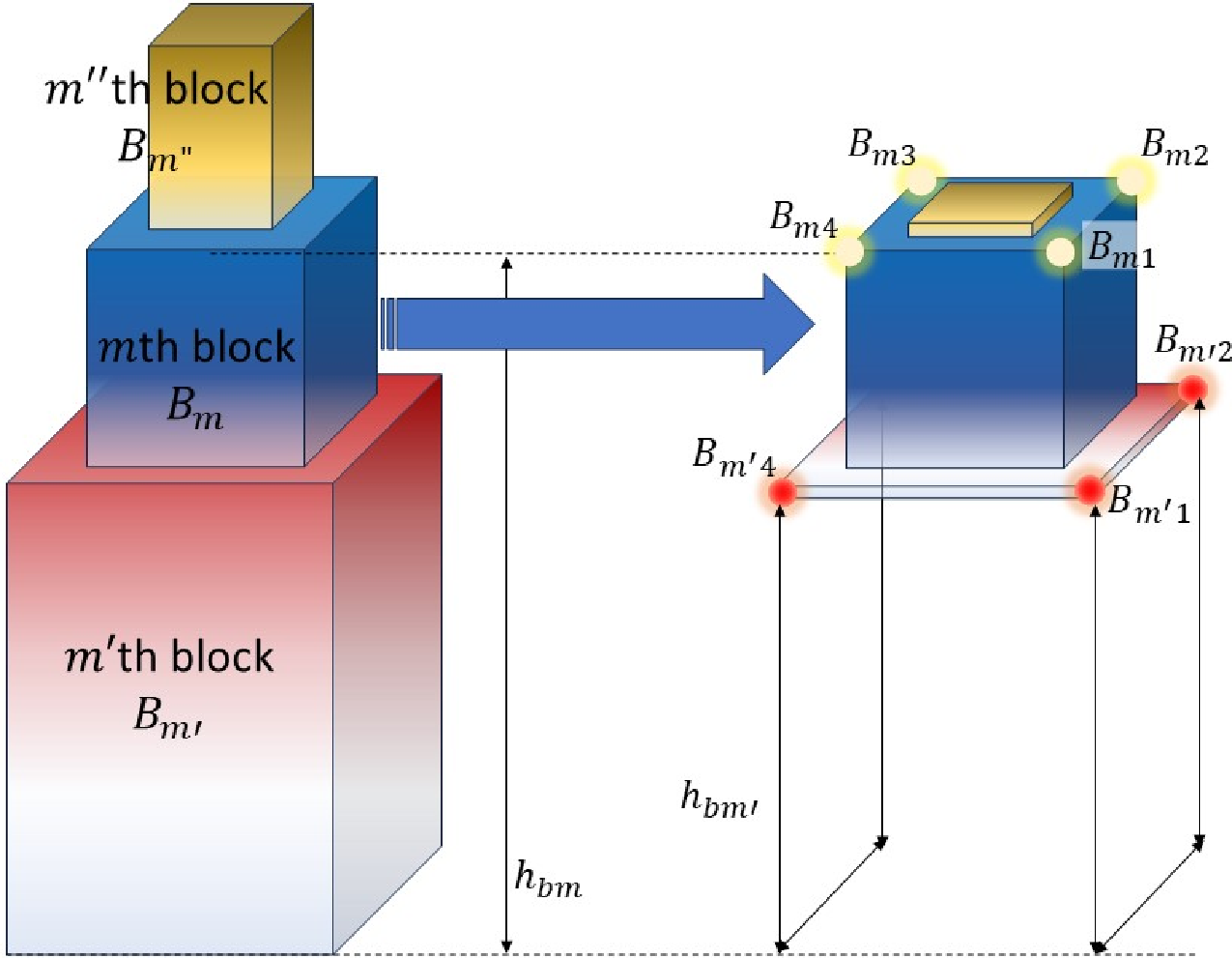}
		\caption{A graphical example of a 3D building block which can be approximated with three cubes, $B_{m}$, $B_{m'}$, and $B_{m"}$. }
		\label{sb1}
	\end{center}
\end{figure}

\subsection{LoS Coverage Model for more General Block Shapes}
Almost most of the building blocks in the cities are either rectangular cubes or can be approximated with rectangular cubes. A graphical example is provided in Fig. \ref{sb1}, where a building block is modeled with three cubes, $B_{m}$, $B_{m'}$, and $B_{m"}$. For this figure, the NLoS coverage generated by $B_{m'}$ is obtained using Algorithm 1. However, to model the NLoS coverage obtained from $B_m$ and $B_{m"}$, it is necessary to make a series of simple modifications in Algorithm 1. For example, we perform these modifications for $B_m$ and it can be applied to the rest of the blocks. Let $S_m$ specify the area between the points $B_{m1}$, $B_{m2}$, $B_{m3}$, and $B_{m4}$, and $S_{m'}$ specify the area between  the points $B_{m'1}$, $B_{m'2}$, $B_{m'3}$, and $B_{m'4}$.
The modifications are made as follows:
\begin{itemize}
\item $h_{bm'}$ should be used instead of $z_{bm'}$ (area surface of $S$) in \eqref{c1}, \eqref{c2}, \eqref{c3}, and \eqref{c4}.
\item In order to obtain the NLoS coverage on the roof of $B_{m'}$, use $h_{ij}$ instead of $h_{m'}$ for the cell grids $s_{ij}$s placed in $S_{m'}$.
\end{itemize}

\begin{figure}
	\begin{center}
		\includegraphics[width=3.4 in]{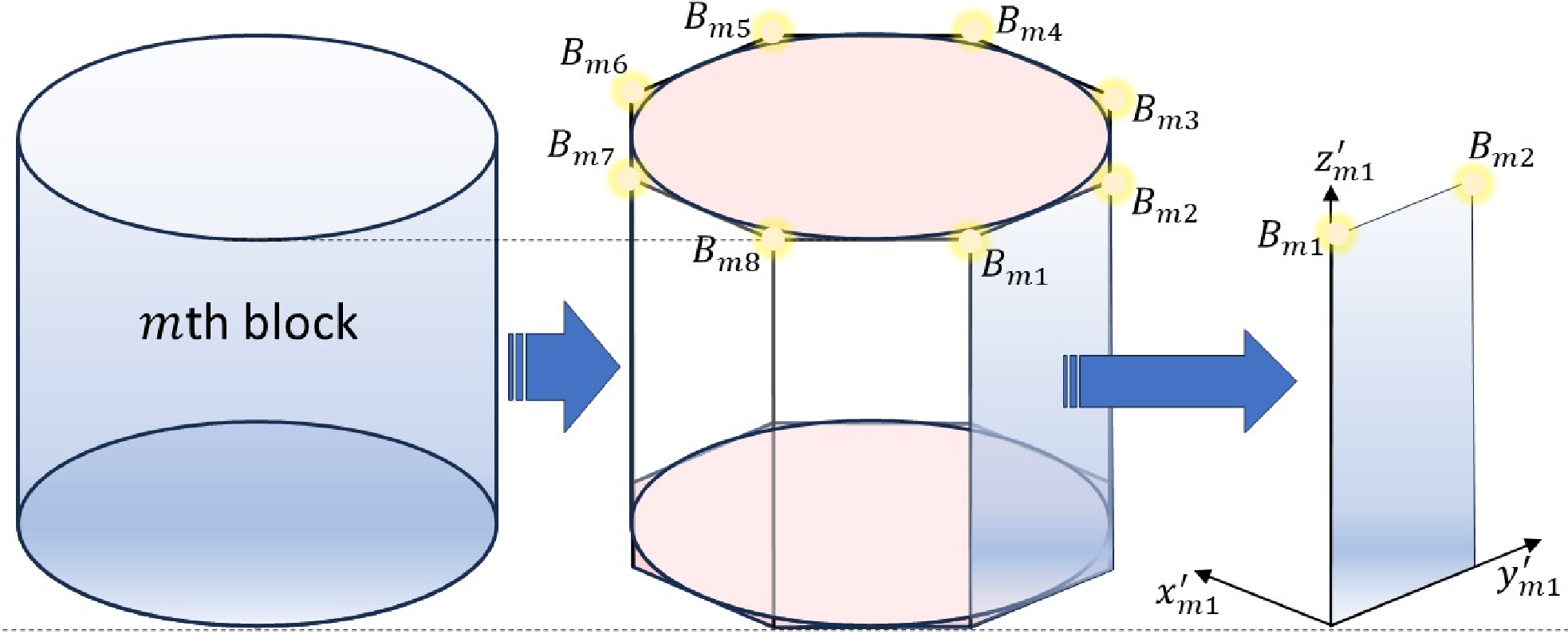}
		\caption{A graphical example of approximating a cylindrical-shaped obstacle with a cylindrical octagon. Then, for each lateral plane, the associated coordinate system is defined to calculate the resulting NLoS coverage by using Algorithm 1.}
		\label{sb2}
	\end{center}
\end{figure}

Another category of 3D obstacles has circular, elliptical, and semi-circular shapes. In this case, elliptical cylinders can be approximated with polygonal cylinders. In the simplest case, an elliptic cylinder can be approximated with a 4-sided cylinder (cube) so that the elliptic cylinder fits in the cube. The coverage obtained in this case is the worst case because the obtained NLoS coverage percentage is slightly more than the actual NLoS coverage percentage. As the number of lateral sides (denoted by $N_\text{side}$) increases, the polygonal cylinder approximation becomes closer to the elliptical cylinder. For example, in Fig. \ref{sb2}, an elliptical cylinder is approximated by an octagonal cylinder. In this case, NLoS coverage can be obtained with some simple modifications in Algorithm 1. Unlike the cube (where all four lateral sides have the same coordinate), here, as shown in Fig. \ref{sb2}, an independent coordinate plane should be created for each lateral plane. Let $[x'_{mi},y'_{mi},z'_{mi}]$ for $i\in\{1,...,N_\text{side}\}$ denote the coordinate plane of each lateral plane. 
Then, using \eqref{s4}, we transfer all the points in the main $[x,y,z]$ coordinate plane to the each coordinate plane $[x'_{mi},y'_{mi},z'_{mi}]$ and obtain the NLoS coverage using Algorithm 1.

\begin{figure}
	\begin{center}
		\includegraphics[width=2.8 in]{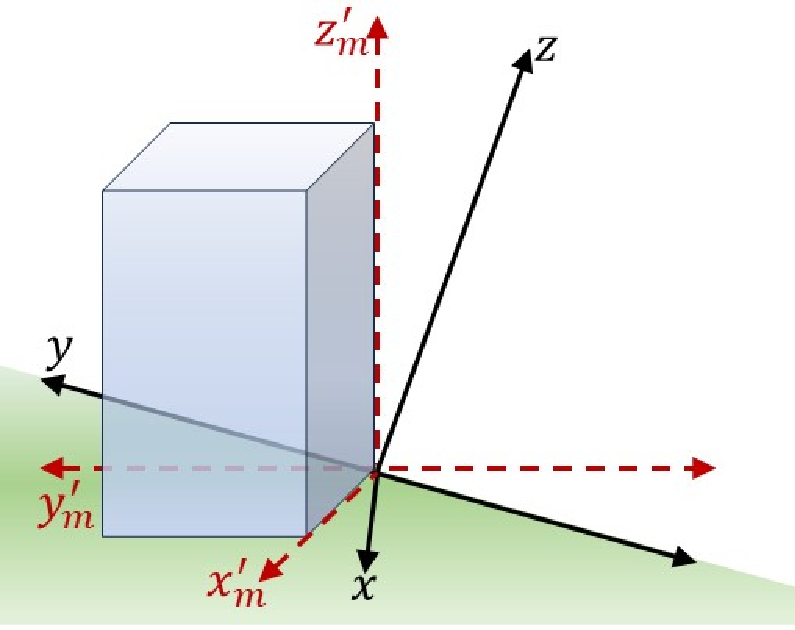}
		\caption{A graphical example when the $z$ axis of the $m$th block is not aligned with the $z$ axis of the main coordinate plane.}
		\label{sb3}
	\end{center}
\end{figure}

Most of the existing obstacles can be approximated with a number of cubes. In some cases, however, the $z$ axis of the main coordinate system is not necessarily in the same direction of the $z'_m$ axis of the block. In this case, \eqref{s4} is no longer valid.
Here, to convert the main coordinates to coordinates of the $m$th block $[x'_m,y'_m.z'_m]$, the main coordinate plane must be respectively rotated as $\theta_{mx}$, $\theta_{my}$, and $\theta_{mx}$ around the $z$, $y$, and $x$ axes to achieve the new coordinate plane $[x'_m,y'_m.z'_m]$. By doing so, \eqref{s4} is modified as \eqref{sd1}. After performing this coordinate plane transformation, the rest of the steps of Algorithm 1 are similar to the simple cubic block case.
\begin{figure*}[!t]
	\normalsize
	\begin{align}
		\label{sd1}
		\underbrace{\left( \begin{matrix} 
				x_f  \\
				y_f   \\
				z_f  \\
			\end{matrix}  \right)}_{{\mathbf{f}}}
		= 
		\underbrace{ \left( \begin{matrix} 
				\cos(\theta_{mz}) & -\sin(\theta_{mz}) & 0  \\
				\sin(\theta_{mz}) & \cos(\theta_{mz}) & 0  \\
				0 & 0 & 1  \\
			\end{matrix}  \right) }_{{\mathbf{R_z}}} 
		%
		\underbrace{ \left( \begin{matrix} 
				\cos(\theta_{my}) & -\sin(\theta_{my}) & 0  \\
				\sin(\theta_{my}) & \cos(\theta_{my}) & 0  \\
				0 & 0 & 1  \\
			\end{matrix}  \right) }_{{\mathbf{R_y}}} 
		%
		\underbrace{ \left( \begin{matrix} 
				\cos(\theta_{mx}) & -\sin(\theta_{mx}) & 0  \\
				\sin(\theta_{mx}) & \cos(\theta_{mx}) & 0  \\
				0 & 0 & 1  \\
			\end{matrix}  \right) }_{{\mathbf{R_x}}} 
		\underbrace{\left( \begin{matrix} 
				x'_f  \\
				y'_f   \\
				z'_f   \\
			\end{matrix}    \right)}_{{\mathbf{f}}'}.
	\end{align}
	\hrulefill
\end{figure*}

\section{Optimal Positioning Based on Definition 1} \label{Sec3}
In this section, we focus on increasing the LoS coverage percentage based on definition 1. Using the general results obtained in this section, in the next section, optimal design and analysis of a THz-based communication system based on definition 2 is investigated to provide LoS coverage for distributed ground nodes between 3D obstacles.

Let $A=d_x\times d_y$ represents the considered area for the positions of UAVs which is an area parallel to $S$ in the sky.
For UAVs' positioning, we first partition the  area $A$ into $N_{ux}\times N_{uy}$ grid cells. Each grid cell is denoted by $a_{ij}=d_{uxi}\times d_{uyj}$ where $d_{uxi}=d_x/N_{ux}$, and $d_{uyj}=d_y/N_{uy}$. Next, we seek to find the set of optimal points in $A$ for positioning the set of UAVs.
\begin{align}
	\label{o1}
	 \underset{ p_1,...,p_{N_u} }
	{\max}
	& & \mathcal{C} =\frac{\mathbb{I}} {N_x\times N_y}\times 100,
\end{align}
where $\mathbb{I}$ is the number of non-zero entries of ${\bf C}=\sum_{n=1}^{N_u} {\bf C}_n$.
Although the optimization problem \eqref{o1} seems simple, the computational load to find the optimal values for the set of $\{p_1,...,p_{N_u}\}$ is very heavy. Note that the number of possible states for a system with only one UAV is equal to $N_{ux}\times N_{uy}$. Therefore, the number of possible states for the set of $\{p_1,...,p_{N_u}\}$ is $(N_{ux}\times N_{uy})^{N_u}$. For example, for a target area $S=500\times500$ m$^2$ and $N_u=5$, and assuming an accuracy of 1 m, the total number of possible states for the set of $\{p_1,...,p_{N_u}\}$ is $(500\times500)^{5}\simeq 10^{27}$. It should be noted that the matrix ${\bf C}=\sum_{n=1}^{N_u} {\bf C}_n$ must be obtained $10^{27}$ times which is itself a $500\times500$ matrix and for each entry of that matrix we should check the spacial 3D vectors introduced in Algorithm 1 at least $4\times M_b$ times where $M_b$ is the number of 3D blocks.

\subsection{Positioning Using Greedy Algorithm}
According to our topology, using the greedy algorithm to solve problem \eqref{o1} can be an appropriate tool. To use the greedy algorithm, it is necessary to first define the set of actions for $u_n$. Let's assume that $u_n$ is located at $p_n(t)$ at time $t$. At time $t+1$, the UAV can perform one of the following 5 allowed actions $a_n(t)$ and go to the position $p_n(t+1)$ as:
\begin{align}
	\label{a1}
	p_n(t+1) = p_n(t) + a_n(t)
\end{align}
where $a_n(t)\in\{0,d_{uxi},-d_{uxi},d_{uyj},-d_{uyj}\}$. In other words, at any moment, the UAV decides whether to stay still, move forward, backward, left, or right one unit.
In the greedy algorithm, we decide to choose the action $a(t)$ as follows:
\begin{align}
	\label{greed1}
	\underset{ a_n(t) }
	{\max}
	& & \mathcal{C}_n =\frac{\mathbb{I}_n[p_n(t)+a_n(t)]} {N_x\times N_y}\times 100,
\end{align}
where $\mathbb{I}_n[p_n(t)+a_n(t)]$ is the number of entries of ${\bf C}_n$ which have non-zero values for $p_n(t+1)=p_n(t)+a_n(t)$.
Note that we have a set of UAVs that decide to collaborate together to increase the LoS coverage percentage. Therefore, we have to modify \eqref{greed1} for the set of UAVs as follows:
\begin{align}
	\label{greed2}
	\underset{{\bf a}(t) }
	{\max}
	& & \mathcal{C} =\frac{\mathbb{I}[{\bf p}(t)+{\bf a}(t)]} {N_x\times N_y}\times 100,
\end{align}
where ${\bf p}(t)=[p_1(t),...,p_{N_u}]$ is a vector of the position of all UAVs at the time $t$, and 
$\mathbb{I}[{\bf p}(t)+{\bf a}(t)]$ is the number of entries of ${\bf C} = \sum_{n=1}^{N_u}{\bf C}_n$ which have non-zero values for $[{\bf p}(t)+{\bf a}(t)]$. 
Also, ${\bf a}(t)\in\{{\bf a}_1(t), ...,{\bf a}_{N_u}(t)\}$ is a vector that contains the action of all the UAVs, where
\begin{align}
	\label{a2}
	{\bf a}_n(t) = [0,...,0,a_n(t),0,...,0].
\end{align}
According to \eqref{greed2} and \eqref{a2}, at each time $t$, only one UAV can move and the locations of the rest of the UAVs are fixed. Therefore, at each time $t$, the best action is selected among $5\times N_u$ possible states. To select optimal action using \eqref{greed2} among $5\times N_u$ possible states, it is necessary to compute $5\times N_u$ times the matrix ${\bf C}_n$  using Algorithm 1.
Based on \eqref{greed2}, the greedy algorithm starts with a random value for the positions of the UAVs and performs a series of actions to reach a local maximum point. At the local maximum point, all UAVs choose $a_n(t)=0$, which means that the position of the UAVs provides a better LoS coverage percentage than the $(5\times N_u-1)$ neighboring points.
To achieve the global optimal point, it is necessary to run the proposed greedy algorithm with a large number of random starting points so that one of the local optimal points is the global optimal point. The optimal positioning based on the greedy algorithm is summarized in Algorithm 2.

\begin{algorithm}
	\caption{Positioning based on Greedy Method}
	\begin{algorithmic}[1]
		\renewcommand{\algorithmicrequire}{\textbf{Input:}}
		\renewcommand{\algorithmicensure}{\textbf{Output:}}
		\REQUIRE $N_\text{g,it}$, $A=d_x\times d_y$, $d_{uxi}$, $d_{uyj}$, $N_{ux}$, $N_{uy}$, $S=d_x\times d_y$, $N_x$, $N_y$, $N_u$, $M_b$, $h_\text{max}$, and information about 3D obstacles
		\ENSURE  ${\bf C}_\text{opt}$, $\mathcal{C}_\text{opt}$, ${\bf p}_{u,\text{opt}}=\{p_{1}, ...,p_{N_u}\}$
		\\ \textit{Initialize ${\bf C}=0$ and $\mathcal{C}=0$}
	\FOR {$I=1 : N_\text{g,it}$}
	\STATE Generate random position for UAVs ${\bf p}=\{p_1,...,p_{N_u}\}$
	\STATE Initialize $B_\text{break}=0$
	\WHILE {$B_\text{break}=0$}
	\FOR {$j=1:5$ and $n=1 : N_u$}
	\STATE Update ${\bf p}'={\bf p} + {\bf a}_{n,j}$
	\STATE Compute $\{ {\bf C}_1,...,{\bf C}_{N_u}\}$ using Algorithm 1
	\STATE Compute ${\bf C}(n,j)$ and then $\mathcal{C}(n,j)$
	\ENDFOR
	\STATE Find $(n,j)$ with the maximum $\mathcal{C}(n,j)$
	\STATE Update ${\bf p}={\bf p} + {\bf a}_{n,j}$
	\\ \textit{\% Check we have reached the local maximum point}
	\IF {${\bf a}_{n,j}=0$}
	\STATE $B_\text{break}=1$
	\ENDIF
	\ENDWHILE
	\\ \textit{\%  Update the optimal values}
	\IF {$\mathcal{C}(n,j)>\mathcal{C}_\text{opt}$}
	\STATE Update $\mathcal{C}_\text{opt}=\mathcal{C}(n,j)$, ${\bf C}_\text{opt}={\bf C}(n,j)$,
	${\bf p}_{u,\text{opt}}={\bf p}$
	\ENDIF
	\ENDFOR
\end{algorithmic} 
\end{algorithm}

Let $N_\text{g,step}(I)$ represent the number of steps taken from the random starting point to reach the local maximum point, and $N_\text{g,it}$  represents the number of iterations until reaching the global maximum point. 
Since the main part of the computational complexity is related to the matrix ${\bf C}_n$, the computational complexity at each time $t$ is proportional to the $\mathcal{O}_\text{gr}(t)\propto 5\times N_u$. Therefore, the computational complexity of Algorithm 2 is proportional to
\begin{align}
	\label{o3}
	\mathcal{O}_\text{gr} \propto 5\times N_u \times \sum_{I=1}^{N_\text{g,it}} N_\text{g,step}(I).
\end{align}

\subsection{Positioning based on Genetic Algorithm}
The initial value is the main challenge of the proposed greedy-based algorithm. If the initial value is not chosen correctly, it will frequently get stuck at local maximum points. To solve this problem, in this section, we propose a GA-based positioning algorithm. For the evolution (moving towards the optimal positions for a set of UAVs), the GA usually starts from a random population (a set of random positioning for UAVs), and iteratively modifies populations by using elite selection, crossover, and mutation.
The GA-based positioning method is provided in Algorithm 3.
Based on Algorithm 3, we create a random collection (initial population) of the set of UAVs in the area $A$ as follows:
\begin{align}
	\label{ga1}
	{\bf P}_\text{pop} = \begin{pmatrix} 
		p_1(1) & p_2(2) & \cdots & p_1(I_p) \\
		p_2(1) & p_2(2) & \cdots & p_2(I_p) \\ 
		\vdots & \vdots & \vdots & \vdots\\ 
		p_{N_u}(1) & p_{N_u}(2) & \cdots & p_{N_u}(I_p) 
	\end{pmatrix},
\end{align}
where each column of ${\bf P}_\text{pop}$ represents a random set for the position of UAVs and $I_p$ represents the population size. 
Using \eqref{greed2} and Algorithm 1, we obtain the LoS coverage percentage for each set of UAV positions.
Let $\mathcal{C}_\text{ga,1}=[\mathcal{C}(1), ...,\mathcal{C}(I_p)]$ denotes the LoS coverage vector corresponding to the initial population ${\bf P}_\text{pop}$.
At each iteration, using $\mathcal{C}_\text{ga,1}$, the GA selects columns from the current population ${\bf P}_\text{pop}$ to be parents and applies crossover and mutation on them to generate new positions ($N_u\times I_\text{cros}$ matrix ${\bf p}_\text{cros}$ related to the crossover and $N_u\times I_\text{mut}$ matrix ${\bf p}_\text{mut}$ related to the mutation).
As elitism, we select $I_\text{elit}$ maximum values of $\mathcal{C}_\text{ga,1}$. 
Let the set $\{m_1, ..., m_{I_\text{elit}} \}$ represents the number of $I_\text{elit}$ maximum values of $\mathcal{C}_\text{ga,1}$. Therefore, in each iteration, the elitism matrix is defined as:
\begin{align}
	\label{ga2}
	{\bf P}_\text{elit} = \begin{pmatrix} 
		p_1(m_1) & p_2(m_2) & \cdots & p_1(m_{I_\text{elit}}) \\
		p_2(m_1) & p_2(m_2) & \cdots & p_2(m_{I_\text{elit}}) \\ 
		\vdots & \vdots & \vdots & \vdots\\ 
		p_{N_u}(m_1) & p_{N_u}(m_2) & \cdots & p_{N_u}(m_{I_\text{elit}}) 
	\end{pmatrix}.
\end{align}
At the end of each iteration, the matrix ${\bf P}_\text{pop}$ is updated as ${\bf p}_\text{pop}=[{\bf p}_\text{elit}, {\bf p}_\text{cros}, {\bf p}_\text{mut}]$. 
In each iteration, the computational complexity of Algorithm 3 is mainly due to the calculation of $\mathcal{C}_\text{ga,1}=[\mathcal{C}(1), ...,\mathcal{C}(I_p)]$ using Algorithm 1 for the population members. Considering that this algorithm is repeated $N_\text{ga,it}$ times, the computational complexity related to the implementation of Algorithm 3 is proportional to 
\begin{align}
	\label{o4}
	\mathcal{O}_\text{ga} \propto I_p \times N_u \times N_\text{ga,it}.
\end{align}

\begin{algorithm}
	\caption{Positioning based on GA Algorithm}
	\begin{algorithmic}[1]
		\renewcommand{\algorithmicrequire}{\textbf{Input:}}
		\renewcommand{\algorithmicensure}{\textbf{Output:}}
		\REQUIRE $I_p$, $I_\text{elit}$, $I_\text{mut}$, $I_\text{cros}$, $N_\text{ga,it}$, $A=d_x\times d_y$, $d_{uxi}$, $d_{uyj}$, $N_{ux}$, $N_{uy}$, $S=d_x\times d_y$, $N_x$, $N_y$, $N_u$, $M_b$, $h_\text{max}$, and information about 3D obstacles
		\ENSURE  ${\bf C}_\text{opt}$, $\mathcal{C}_\text{opt}$, ${\bf p}_{u,\text{opt}}=\{p_{1}, ...,p_{N_u}\}$
		\\ \textit{Initialize ${\bf C}=0$ and $\mathcal{C}=0$}
	\STATE Generate initial population, ${\bf p}_\text{pop}$
	\FOR {$I=1 : N_\text{ga,it}$}
	\STATE Compute $\mathcal{C}_\text{ga,1}=[\mathcal{C}(1), ...,\mathcal{C}(I_p)]$ using \eqref{greed2} and Algorithm 1
	\\ \textit{\% Elitism}
	\STATE Find the $I_\text{elit}$ maximum values of $\mathcal{C}_\text{ga,1}$ and put the corresponding positions in ${\bf p}_\text{elit}$
	\\ \textit{\% Crossover and Mutation}
	\STATE Using $\mathcal{C}_\text{ga,1}$, generate parents from ${\bf p}_\text{pop}$ 
	\STATE Roulette Selection of parents
	\STATE Perform crossover to generate new positions in ${\bf p}_\text{cros}$
	\STATE Perform mutation to generate new positions in ${\bf p}_\text{mut}$
	\STATE Update ${\bf p}_\text{pop}=[{\bf p}_\text{elit}, {\bf p}_\text{cros}, {\bf p}_\text{mut}]$
	\ENDFOR
	\\ \textit{\%  Optimal values}
	\STATE Select first column of ${\bf p}_\text{pop}$ as ${\bf p}_{u,\text{opt}}$
	\STATE Compute ${\bf C}_\text{opt}$ using ${\bf p}_\text{opt}$ and Algorithm 1
	\STATE Compute $\mathcal{C}_\text{opt}$ using ${\bf C}_\text{opt}$ and \eqref{greed2}
\end{algorithmic} 
\end{algorithm}

\section{Hybrid GA-Greedy Positioning}
The main problem of the GA-based algorithm is that it converges slowly. On the other hand, the main challenge of the greedy-based algorithm is the random initial value, which causes it to get stuck in local maximum values. Using GA and greedy algorithms, a hybrid GA-Greedy positioning method is proposed in Algorithm 4. 
Similar to the GA algorithm, Algorithm 4 generates a random initial population for UAVs' positions. 
Then, it iteratively tries to move toward the optimal position for UAVs. To this end, in each iteration, similar to the GA, it first applies the operations of elitism, crossover, and mutation on ${\bf p}_\text{pop}$ to achieve a new update of ${\bf p}_\text{pop}$.
Then, it uses the elite values of ${\bf p}_\text{pop}$ as the initial values for the Greedy algorithm. In other words, the proposed hybrid method overcomes the weakness of the greedy algorithm in the initial random value and makes it converge toward the optimal position faster.

\begin{algorithm}
	\caption{Hybrid GA-Greedy Algorithm}
	\begin{algorithmic}[1]
		\renewcommand{\algorithmicrequire}{\textbf{Input:}}
		\renewcommand{\algorithmicensure}{\textbf{Output:}}
		\REQUIRE $N'_\text{greed}$, $I_\text{greed}$, $I_p$, $I_\text{elit}$, $I_\text{mut}$, $I_\text{cros}$, $N_\text{ga,it}$, $A=d_x\times d_y$, $d_{uxi}$, $d_{uyj}$, $N_{ux}$, $N_{uy}$, $S=d_x\times d_y$, $N_x$, $N_y$, $N_u$, $M_b$, $h_\text{max}$, and information about 3D obstacles
		\ENSURE  ${\bf C}_\text{opt}$, $\mathcal{C}_\text{opt}$, ${\bf p}_{u,\text{opt}}=\{p_{1}, ...,p_{N_u}\}$
		\\ \textit{Initialize ${\bf C}=0$ and $\mathcal{C}=0$}
	\STATE Generate initial population, ${\bf p}_\text{pop}$
	\FOR {$I=1 : N_\text{ga,it}$}
	\STATE Compute $\mathcal{C}_\text{ga,1}=[\mathcal{C}(1), ...,\mathcal{C}(I_p)]$ using \eqref{greed2} and Algorithm 1
	\\ \textit{\% Elitism}
	\STATE Find the $I_\text{elit}$ maximum values of $\mathcal{C}_\text{ga,1}$ and put the corresponding positions in ${\bf p}_\text{elit}$
	\\ \textit{\% Crossover and Mutation}
	\STATE Using $\mathcal{C}_\text{ga,1}$, generate parents from ${\bf p}_\text{pop}$ 
	\STATE Roulette Selection of parents
	\STATE Perform crossover to generate new positions in ${\bf p}_\text{cros}$
	\STATE Perform mutation to generate new positions in ${\bf p}_\text{mut}$
	\STATE Update ${\bf p}_\text{pop}=[{\bf p}_\text{elit}, {\bf p}_\text{cros}, {\bf p}_\text{mut}]$
	\FOR {$I'=1 : I_\text{greed}$}
	\STATE Random selection of one column from the first $N'_\text{greed}$ columns of ${\bf p}_\text{pop}$ as ${\bf p}$ 
	\\ \textit{\% Run the greedy algorithm using the desired GA outputs}
	\STATE Initialize $B_\text{break}=0$
	\WHILE {$B_\text{break}=0$}
	\FOR {$j=1:5$ and $n=1 : N_u$}
	\STATE Update ${\bf p}'={\bf p} + {\bf a}_{n,j}$
	\STATE Compute $\{ {\bf C}_1,...,{\bf C}_{N_u}\}$ using Algorithm 1
	\STATE Compute ${\bf C}(n,j)$ and then $\mathcal{C}(n,j)$
	\ENDFOR
	\STATE Find $(n,j)$ with the maximum $\mathcal{C}(n,j)$
	\STATE Update ${\bf p}={\bf p} + {\bf a}_{n,j}$
	\\ \textit{\% Check we have reached the local maximum point}
	\IF {${\bf a}_{n,j}=0$}
	\STATE $B_\text{break}=1$
	\ENDIF
	\ENDWHILE
	\ENDFOR
	\ENDFOR
	\\ \textit{\%  Optimal values}
	\STATE Select first column of ${\bf p}_\text{pop}$ as ${\bf p}_{u,\text{opt}}$
	\STATE Compute ${\bf C}_\text{opt}$ using ${\bf p}_{u,\text{opt}}$ and Algorithm 1
	\STATE Compute $\mathcal{C}_\text{opt}$ using ${\bf C}_\text{opt}$ and \eqref{greed2}
\end{algorithmic} 
\end{algorithm}

The main problem in the proposed hybrid method using elite values is that in some cases the algorithm may get stuck in a local maximum point. Therefore, in Algorithm 4, instead of elitism values, we randomly select $I_\text{greed}$ values out of $N'_\text{greed}$ maximum values of $\{ {\bf C}_1,...,{\bf C}_{N_u}\}$. In this case, the $N'_\text{greed}$ maximum values include $I_\text{elit}$ values of elitism and $N'_\text{greed}-I_\text{elit}$ values of crossover, whose values change in each iteration.

In each iteration, the computational complexity of Algorithm 4 is proportional to the computational complexity of the GA in each iteration plus the computational complexity of the greedy algorithm for $I_\text{greed}$ iterations. Therefore, the computational complexity of Algorithm 4 after $N_\text{ga,it}$ iterations is obtained as:
\begin{align}
	\label{05}
	\mathcal{O}_\text{ga-gr} \propto N_\text{ga,it} \left( I_p\times N_u  + 5\times N_u \times \sum_{I=1}^{I_\text{greed}} N_\text{g,step}(I)\right).
\end{align}

\begin{figure}
	\begin{center}
		\includegraphics[width=3.4 in]{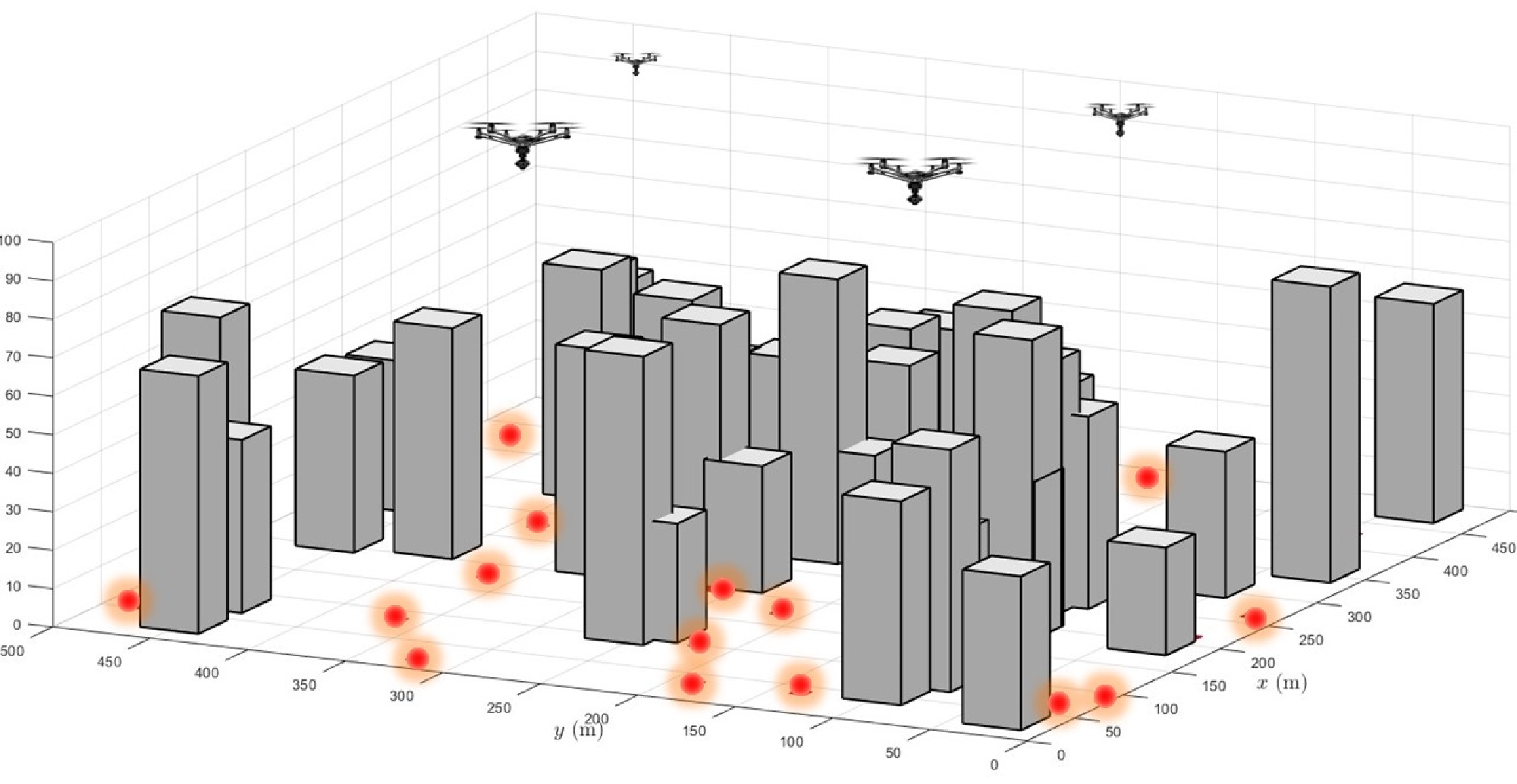}
		\caption{A graphical example of the considered UAV-based THz network to find the optimal position for a set of UAVs to provide LoS THz connections for randomly distributed nodes among the 3D obstacles.}
		\label{sf1}
	\end{center}
\end{figure}

\section{Positioning for THz Wireless networks under 3D blocks}

In the previous section, we examined the problem of UAVs' positioning in an environment with 3D obstacles to maximize the LoS coverage percentage in a target area.
In this section, we examine the problem of UAVs' positioning for a wireless network. As shown in Fig. \ref{sf1}, we assume that a set of nodes is randomly distributed among the 3D obstacles. Our target is to locate a set of UAVs in 3D space in coordination with each other so that the average network capacity is maximized while all the distributed nodes are in LoS state. It should be noted that unlike lower frequency or omnidirectional short-range THz communications (for instance indoor THz network), the LoS state is a necessary condition because high-frequency directional THz links cannot serve nodes behind large buildings. In this case, the optimization problem is formulated as follows:
\begin{subequations}\label{optim1}
	\begin{align}
		\label{opt1}
		& \underset{  p_1,...,p_{N_u}, \mathcal{L}_1,...,\mathcal{L}_{N_u} }
		{\max}
		& &  \mathbb{C} = \frac{1}{N_g}\sum_{n=1}^{N_u} \sum_{k\in \mathcal{L}_n} \log_2\left( 1 + {P_t h_{nk}^2}/{N_0} \right) \\
		\label{opt2}
		& ~~~~~~~~~~~~~~\text{s.t.}   & & \sum_{k\in \mathcal{L}_n} k=N_u,\\
		\label{opt3}
		&& & \mathcal{L}_1\bigcap \mathcal{L}_2   \bigcap ... \bigcap\mathcal{L}_{N_u} = \emptyset, \\
		\label{opt4}
		&&& \mathbb{S}_k=1,~~~k\in\{1,...,N_u\},
	\end{align}
\end{subequations}
where $\mathcal{L}_n$ represents the $n$th cluster or the set of $g_k$ connected to the $n$th UAV, and $\mathbb{S}_k$ represents the status of the $g_k$ in such a way that if $\mathbb{S}_k=1$, the $g_k$ is in LoS state and if $\mathbb{S}_k= 0$, the $g_k$ is in the NLoS state.
In optimization problem \eqref{optim1}, the target is to find optimal positioning for the set of UAVs along with the clustering of the ground nodes by considering the 3D obstacles in such a way that the average capacity of the network is maximized. Constraints \eqref{opt2} and \eqref{opt3} are related to the clustering of the ground nodes and constraint \eqref{opt4} is related to placing all nodes in the LoS state.

\subsection{Greedy, GA, and Hybrid Greedy-GA Algorithms}
Let vector ${\bf p}_{g}=[p_{g1},...,p_{gN_g}]$ and ${\bf p}_u=[p_1,...,p_{N_u}]$ represent the set of positions of $g_{k}$ and $u_n$, respectively..
In the previous optimization problem, the computational complexity is related to obtaining $N_x\times N_y$ matrix ${\bf C}_n$ for all the grid cells in the target area $S$. To compute ${\bf C}_n$, we use the proposed Algorithm 1.
However, for optimization problem \eqref{optim1}, we only need to compute ${\bf C}$ for the vector ${\bf p}_{g}$. To show the distinction, here, we use the notation ${\bf c}_k$ instead of ${\bf C}_n$. 

{\bf Remark 3.} {\it To obtain the vector ${\bf c}_k$, we use Algorithm 1 again, with the difference that instead of the set of $N_x\times N_y$ grid cells in the target area $S$, we use $g_k$ and vector ${\bf p}_{n}$.}

\begin{algorithm}
	\caption{Clustering for any given ${\bf p}_u$}
	\begin{algorithmic}[1]
		\renewcommand{\algorithmicrequire}{\textbf{Input:}}
		\renewcommand{\algorithmicensure}{\textbf{Output:}}
		\REQUIRE ${\bf p}_g$, ${\bf p}_u$, $f_t$, $N_u$, $N_g$, $M_b$, and information about 3D obstacles
		\ENSURE  Clusters $\{\mathcal{L}_1,...,\mathcal{L}_{N_u}\}$ and related average network capacity $\mathbb{C}$
	\FOR {$k=1:N_g$}
	\STATE Compute ${\bf c}_k$ using Algorithm 1
	\STATE Compute ${\bf h}_k$ using $p_k$ and ${\bf p}_u$
	\STATE Compute $\mathbb{C}'={\bf c}_k\odot \log_2\left( 1 + {P_t {\bf h}_k\odot {\bf h}_k}/{N_0} \right)$
    \IF {$\max(\mathbb{C}')>0$} 
    \STATE Find $u_n$ that achieves the highest capacity and add $g_k$ to cluster $\mathcal{L}_n$ \\
    \STATE Update average capacity $\mathbb{C}=\mathbb{C} + \max(\mathbb{C}'/N_g)$
    \ELSE
    \STATE Failed and $g_k$ is in NLoS of all $p_n$s
    \ENDIF
    %
    \ENDFOR 
\end{algorithmic} 
\end{algorithm}

For optimization problem \eqref{optim1}, we cannot directly use Algorithms 2-4 because the target is to find the optimal positioning of UAVs along with the optimal clustering of distributed nodes. 
In order to modify Algorithms 2-4 for optimization problem \eqref{optim1}, Algorithm 5 is provided. For each $g_k$, it provides a clustering of UAVs ${\bf p}_u$ to maximize the capacity while placing most UAVs in the LoS state. In particular, in Algorithm 5, for each $g_k$, it obtains the vector ${\bf c}_k$ using Algorithm 1, which determines which UAVs are in the LoS state. Then, by computing the distance, it obtains the vector capacity $\mathbb{C}'_k$ for the set of UAVs. Finally, in lines 6-11, it performs clustering based on capacity. It should be noted that if all UAVs are in the NLoS state, ${\bf c}_k$ is an all-zero vector and the capacity becomes zero. In this case, $g_k$ is not included in any clustering and it shows that ${\bf p}_u$ is not suitable.

Now, to modify Algorithms 2-4 to solve optimization problem \eqref{optim1}, it is necessary to add two steps. First, for each obtained ${\bf p}_u$, clustering and average capacity using Algorithm 5 are obtained. Second, the target is no longer to maximize the LoS coverage percentage $\mathcal{C}$, but rather to locate the set ${\bf p}_g$ in the LoS state and to maximize the average capacity. Therefore, in order to find the optimal location for the ${\bf p}_u$, the metric capacity $\mathbb{C}$ should be used instead of $\mathcal{C}$.

\subsection{Geometrical-based Clustering and Positioning}
The most important challenge of the modified Algorithms 2-4 to solve optimization problem \eqref{optim1} is that these algorithms get stuck in line 9 of Algorithm 5 in most iterations and fail. In other words, for most iterations, there is no clustering that can cover all ${\bf p}_g$.

\begin{algorithm}
	\caption{Geometrical-based Clustering and Positioning for any given ${\bf p}_u$}
	\begin{algorithmic}[1]
		\renewcommand{\algorithmicrequire}{\textbf{Input:}}
		\renewcommand{\algorithmicensure}{\textbf{Output:}}
		\REQUIRE ${\bf p}_g$, ${\bf p}_u$, $f_t$, $N_u$, $N_g$, $M_b$, and information about 3D obstacles
		\ENSURE  UAVs' positions ${\bf p}_u=[p_1,...,p_{N_u}]$, clusters $\{\mathcal{L}_1,...,\mathcal{L}_{N_u}\}$ and related average network capacity $\mathbb{C}$
	\FOR {$k=1:N_g$}
	\STATE Do lines 2-4 of Algorithm 5
	\IF {$\max(\mathbb{C}')>0$} 
	\STATE  Do lines 6-7 of Algorithm 5
	\ELSE
	\STATE Add $g_k$ in the cluster of the nearest $u_n$
	\ENDIF
	%
	\ENDFOR 
	\\ \textit{\% Update ${\bf p}_u$ using geometrical features of environment }
	\FOR {$n=1 : N_u$}
	\STATE Obtain $\mathcal{A}_n$ using $\mathcal{L}_n$ and Algorithm 1
	\IF {$\mathcal{A}_n\neq \empty$}
	\STATE Update $p_n$ with random point in $\mathcal{A}_n$, then, update $\mathbb{C}_n=  \sum_{k\in \mathcal{L}_n} \log_2\left( 1 + {P_t h_{nk}^2}/{N_0} \right)$
	\ELSE 
	\STATE The current clustering failed
	\ENDIF
	\ENDFOR
	\STATE Update $\mathbb{C} = \frac{1}{N_g}\sum_{n=1}^{N_u} \mathbb{C}_n$
\end{algorithmic} 
\end{algorithm}

To solve this challenge and to increase the speed of convergence of the optimization problem \eqref{optim1}, we propose Algorithm 6 which uses the geometrical features of 3D obstacles and significantly reduces the number of failed iterations in Line 9 of Algorithm 5.
Remember that the geometrical characteristics of 3D obstacles are processed in Algorithm 1. For each ${\bf p}_u$, Algorithm 1 obtains the set of points between the 3D obstacles that are placed in the LoS state. However, we now have ${\bf p}_g$, and the target is to find  ${\bf p}_u$ such that all the nodes have LoS connectivity.
Therefore, we must first make some changes in Algorithm 1 to prepare it for use in algorithm 6.
First, we should use cells $p_{aij}=[x_{ij},y_{ij},h_u]$ in plane $A$ instead of cells $p_{aij}=[x_{ij},y_{ij},h_u]$ in plane $S$. Also, instead of vector ${\bf p}_u$, we should use the position of the nodes in cluster $\mathcal{L}_n$.
Then, using the nodes in cluster $\mathcal{L}_n$ and the characteristics of the 3D obstacles, we obtain 3D space vectors to determine the LoS states for the $N_{ux}\times N_{uy}$ different values of $p_{aij}=[x_{ij},y_{ij},h_u]$ as output of Algorithm 1. In other words, the modified Algorithm 1 obtains the acceptable area in $A$ for $u_n$ wherein the UAV can provide LoS connectivity for all the nodes in cluster $\mathcal{L}_n$.
Let $\mathcal{A}_n$ denote the set of the considered acceptable area in $A$. In Algorithm 6, for each given ${\bf p}_n$, it first performs an initial clustering similar to Algorithm 5, with the difference that in line 6, it adds the NLoS nodes to the closest cluster (closest UAV). 
Then, as described, using the modified Algorithm 1, we obtain $\mathcal{A}_n$ in line 10 of Algorithm 6, and then, we transfer $p_n$ to a random position in $\mathcal{A}_n$ in line 12.

{\bf Remark 4.} {\it The important advantage of Algorithm 6 compared to Algorithm 5 is that it uses the characteristics of 3D obstacles for clustering and positioning, and as a result, it significantly reduces the number of failed iterations of Algorithm 5.}

{\bf Remark 5.} {\it Similar to Algorithm 5, Algorithm 6 is just an iteration and for a vector ${\bf p}_u$, it obtains a clustering and modifies ${\bf p}_u$. Then, the output of Algorithm 6, ${\bf p}_u$, is used as input in one of Algorithms 2-4, and new vector ${\bf p}_u$ is given to 6 as input. By repeating this process, we move towards the optimal solution for problem 12.}

\begin{figure}
	\centering
	\subfloat[] {\includegraphics[width=3.2 in]{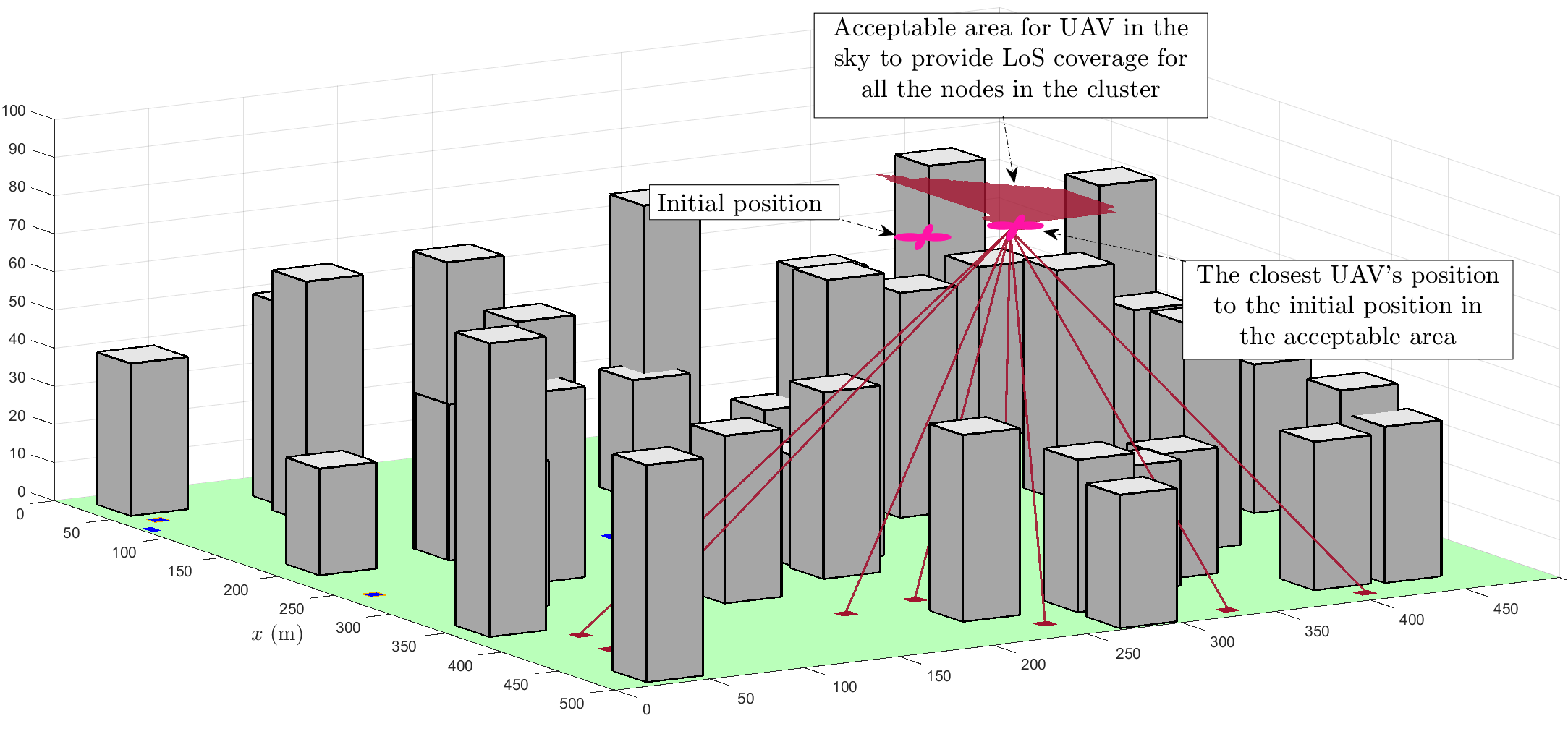}
		\label{cr1}
	}
	\hfill
	\subfloat[] {\includegraphics[width=3.2 in]{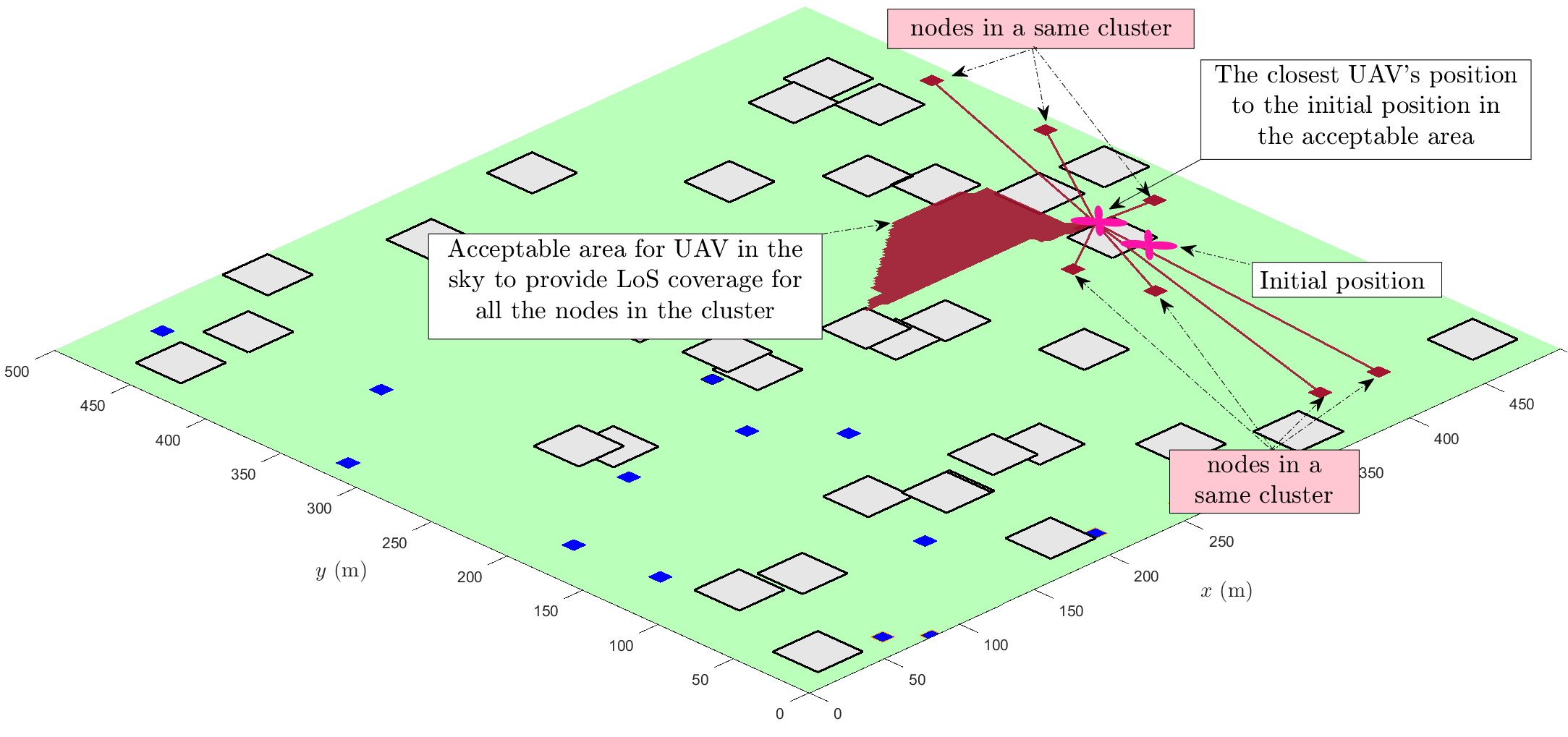}
		\label{cr2}
	}
	\caption{
		An illustration of a clustering of the 25 distributed nodes along with the acceptable area for UAV $\mathcal{A}_n$ obtained by an iteration of Algorithm 7: (a) 3D illustration, and (b) the top view of the target environment.
	}
	\label{cr}
\end{figure}
%

\subsection{Geometrical-K-Means Clustering and Positioning}
As shown in the simulations, Algorithm 6 has a good performance. In the following, we propose a sub-optimal Algorithm 7, which, although it may not reach the optimal solution, it can find a sub-optimal solution for the problem in a very short time. In fact, Algorithm 7 is suitable for applications and networks in which the topology of the network is changing at a faster rate. Therefore, it is necessary to quickly modify the clustering and positioning of UAVs according to the changes in the network topology.

\begin{algorithm}
	\caption{Geometrical-K-Means Clustering and Positioning}
	\begin{algorithmic}[1]
		\renewcommand{\algorithmicrequire}{\textbf{Input:}}
		\renewcommand{\algorithmicensure}{\textbf{Output:}}
		\REQUIRE ${\bf p}_g$, $f_t$, $N_u$, $N_g$, $M_b$, and information about 3D obstacles
		\ENSURE  Optimal UAVs' positions ${\bf p}_{u,\text{opt}}=\{p_{1}, ...,p_{N_u}\}$, clusters $\mathcal{L}_\text{opt}=\{\mathcal{L}_1,...,\mathcal{L}_{N_u}\}$ and maximum average network capacity $\mathbb{C}_\text{max}$
	\\ \textit{ Initialize $\mathbb{C}_\text{max}=0$ }
	\FOR {$I=1:I_\text{it}$}
	\STATE Generate random ${\bf p}_u$ and initialize $B=0$
	\WHILE {$B=0$}
	\STATE ${\bf p}'_u={\bf p}_u$
	\STATE Apply k-means algorithm to obtain new clusters $\{\mathcal{L}_1,...,\mathcal{L}_{N_u}\}$ and ${\bf p}'_u$
	\STATE If ${\bf p}_u={\bf p}_u'$, then update $B=1$ to break WHILE loop, else, update ${\bf p}_u={\bf p}_u'$
	%
	\ENDWHILE
	\\ \textit{\% Update ${\bf p}_u$ using geometrical features of environment }
	\FOR {$n=1 : N_u$}
	\STATE Obtain $\mathcal{A}_n$ using $\mathcal{L}_n$ and Algorithm 1
	\IF {$\mathcal{A}_n\neq \empty$}
	\STATE Update $p_n$ with closest point in $\mathcal{A}_n$, then, update $\mathbb{C}_n=  \sum_{k\in \mathcal{L}_n} \log_2\left( 1 + {P_t h_{nk}^2}/{N_0} \right)$
	\ELSE 
	\STATE The current clustering failed
	\ENDIF
	\ENDFOR
	\\ \textit{\%  Update the optimal values}
	\IF {$\mathbb{C}_\text{max} <\frac{1}{N_g}\sum_{n=1}^{N_u} \mathbb{C}_n$}
	\STATE Update $\mathbb{C}_\text{max} = \frac{1}{N_g}\sum_{n=1}^{N_u} \mathbb{C}_n$, 
	$\mathcal{L}_\text{opt}=\{\mathcal{L}_1,...,\mathcal{L}_{N_u}\}$,
	${\bf p}_{u,\text{opt}}={\bf p}_u$
	\ENDIF
	%
	\ENDFOR 
\end{algorithmic} 
\end{algorithm}

\section{simulation Results}

\begin{figure*}
	\centering
	\subfloat[] {\includegraphics[width=3.4 in]{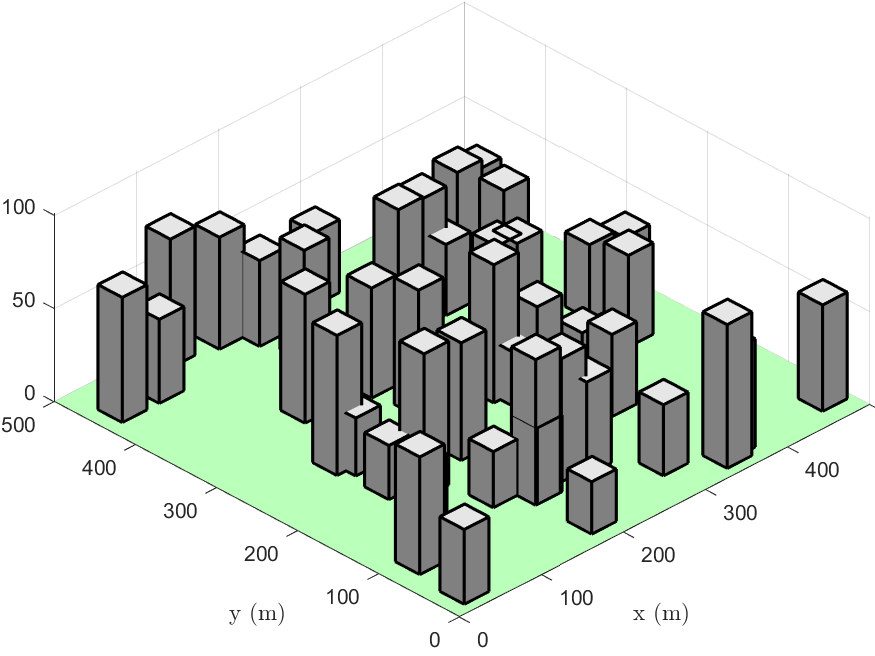}
		\label{cf1}
	}
	\hfill
	\subfloat[] {\includegraphics[width=3.4 in]{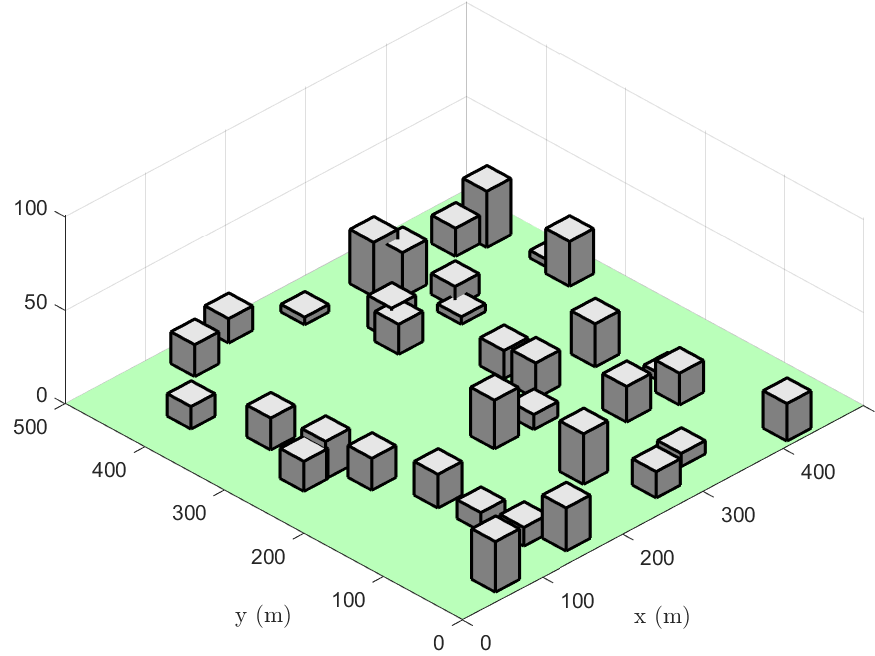}
		\label{cf2}
	}
	\caption{Two different $500\times 500$ m$^2$ environments with 3D obstacles used for simulations including: (a) 45 blocks with an average height of 40 m as an example of an urban environment, and (b) 35 blocks with an average height of 12 m as an example of an sub-urban environment.
	}
	\label{cf}
\end{figure*}
%

The idea used in Algorithm 6 is that in each iteration, we first generate a random distribution of ${\bf p}_u$. Then, without considering the obstacles, we use the k-means algorithm to cluster the distributed nodes \cite{likas2003global}. 
The k-means algorithm itself is an iterative algorithm that, by clustering nodes based on the square of the distance, places the position of the UAV in the average position of the nodes of each cluster \cite{likas2003global}.
However, in the presence of 3D obstacles, k-means clustering is not a suitable solution here because there is no guarantee that all nodes of a cluster are in LoS state. Therefore, in lines 8-15 of Algorithm 7, we obtain the set of $\mathcal{A}_n$ using the characteristics of 3D obstacles (similar to Algorithm 6). Then, we move the position of the UAV obtained from the k-means algorithm to the nearest point in $\mathcal{A}_n$.
To get a better view of Algorithm 7, an iteration of Algorithm 7 is provided in Fig. \ref{cr}. In this example, 25 nodes are randomly distributed among the 3D obstacles. The results show both the 3D environment with 3D obstacles and a top view of the target environment. First, a clustering of the environment is done using the k-means algorithm. In this figure, only one cluster is shown with red nodes along with the corresponding UAV. The k-means algorithm obtains an initial position for the UAV without considering 3D obstacles. Then, using the characterizations of the 3D obstacles, by using Algorithm 1, it obtains the acceptable area for UAV $\mathcal{A}_n$. By moving the UAV in the $\mathcal{A}_n$, all nodes of the considered cluster will be placed in the LoS state. 
Since the initial position of the UAV creates the highest average capacity for the cluster regardless of the NLoS state, therefore, to reach a capacity close to the initial position, we choose the closest point in $\mathcal{A}_n$ for the UAV. As can be seen, in the new position, the UAV creates LoS coverage for all nodes in the cluster.

Although, with each random initial ${\bf p}_u$, the k-means algorithm achieves a different clustering, however, the number of different clusters is limited and therefore, Algorithm 7 converges quickly.

The simulations are divided into two parts related to LoS coverage based on definition 1 and also LoS coverage for a network with a random distribution of nodes based on definition 2.

\subsection{UAVs' Positioning to  Increase LoS coverage Percentage}
To investigate the LoS coverage percentage, two different environments with 3D obstacles are considered, which are shown in Figs. \ref{cf1} and \ref{cf2}. In Fig. \ref{cf1}, 45 blocks with an average height of 40 meters are randomly distributed in a $500\times 500$ m$^2$. In Fig. \ref{cf2}, 35 blocks with an average height of 12 meters are distributed in a $500\times 500$ m$^2$, i.e., $d_x=d_y=500$ m. Also, to obtain the LoS coverage percentage, we divide the area $S=500\times 500$ m$^2$ and $A=500\times 500$ m$^2$ into 1 m$^2$ grid cells, i.e., $d_{xi}=d_{yj}=d_{uxi}=d_{uyj}=1$ m.

\begin{figure}
	\begin{center}
		\includegraphics[width=3.4 in]{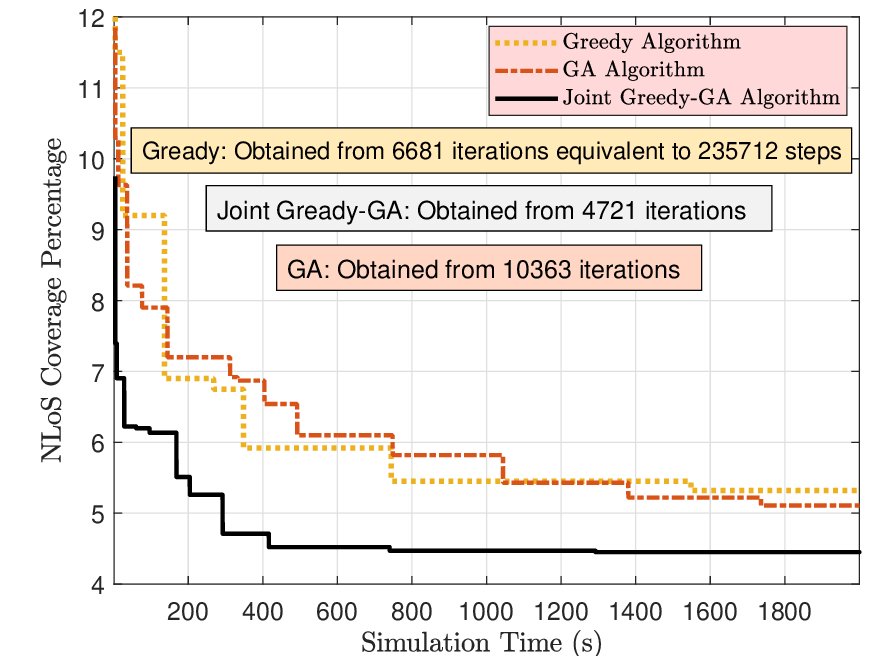}
		\caption{Comparison of the NLoS coverage percentage obtained by Algorithms 2-4 versus running time.}
		\label{bm0}
	\end{center}
\end{figure}
%

First, the performance of different algorithms is investigated in Fig. \ref{bm0} to achieve the lowest NLoS coverage percentage. The results of Fig. \ref{bm0} are obtained for four UAVs at an altitude of 100 m above the environment shown in Fig. \ref{cf1}. These four UAVs use the different methods provided in Algorithms 2-4 and try to adjust their positions in coordination with each other above the 3D obstacles in such a way as to reach the lowest NLoS coverage percentage. As discussed, algorithms 2-4, have different computational complexity in each iteration. Therefore, for a better comparison of the convergence speed of these algorithms, the results are provided versus time.
As can be seen, the combined greedy-GA positioning method converges to the optimal position faster. In fact, this algorithm solves the problem related to the initial value of the greedy algorithm by using the optimal positions obtained from the GA algorithm.

\begin{figure}
	\begin{center}
		\includegraphics[width=3.4 in]{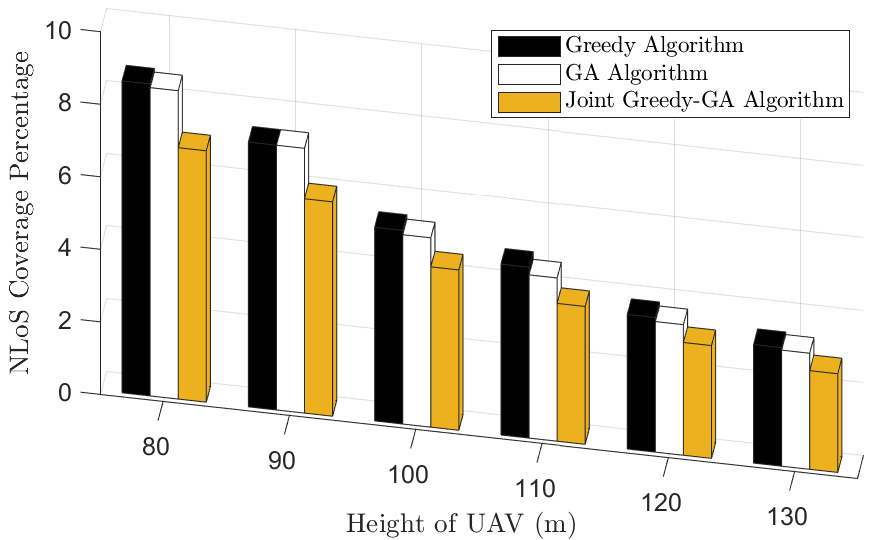}
		\caption{NLoS coverage percentage for different UAVs' heights. }
		\label{bm1}
	\end{center}
\end{figure}
%
\begin{figure*}
	\centering
	\subfloat[] {\includegraphics[width=3.4 in]{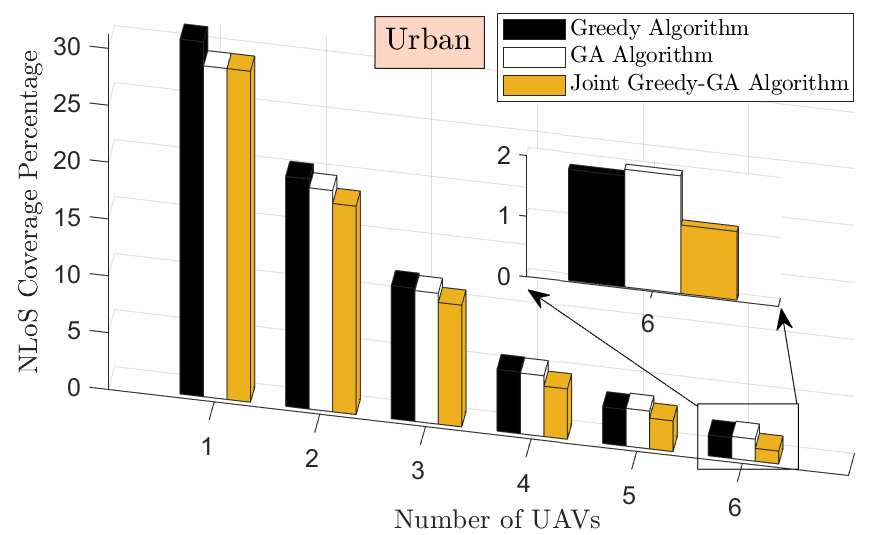}
		\label{ck1}
	}
	\hfill
	\subfloat[] {\includegraphics[width=3.4 in]{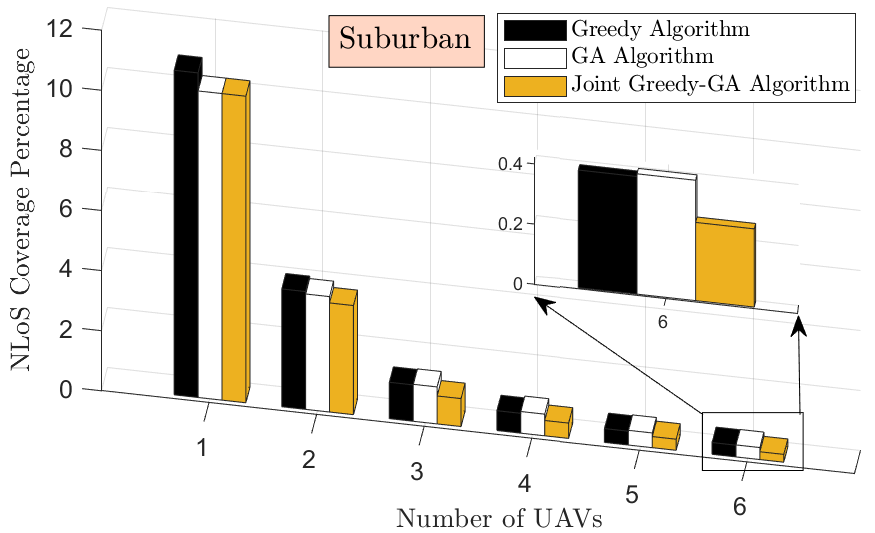}
		\label{ck2}
	}
	\caption{Effect of UAVs' number on NLoS coverage percentage for (a) the 3D environment of Fig. \ref{cf1}, and (b) the 3D environment of Fig. \ref{cf2}.  
	}
	\label{ck}
\end{figure*}
%

One of the important parameters affecting the LoS coverage percentage is the UAVs' height. To this end, in Fig. \ref{bm1}, the LoS coverage percentage is plotted for a wide range of UAVs' heights. The results are obtained for four UAVs. As expected, with the increase in the height of the UAVs, the LoS coverage percentage improves. However, it should be noted that the height of UAVs is usually limited by the rules related to flight security and safety. Another point is that, taking into account wireless communication metrics such as network capacity (which is discussed in the next section) or the outage probability, the network performance does not necessarily improve with the increase in height, and the optimal height selection requires more detailed analyses.

In Fig. \ref{ck}, the effect of the number of UAVs on the NLoS coverage percentage is studied. The results of Figs. \ref{ck1} and \ref{ck2} are obtained by using the optimal positioning of the UAVs at a height of 100 m above the areas of Figs. \ref{cf1} and \ref{cf2}, respectively.
 As expected, by increasing in the number of UAVs, the NLoS coverage percentage decreases. In addition, with the increase in the number of UAVs, the importance of using the proposed hybrid greedy-GA algorithm increases. For example, for $N_u= 2$, based on the results of Fig. \ref{ck1}, we achieve the NLoS  coverage percentages of 20.2, 19.2, and 18.3 for greedy, genetic, and hybrid greedy-GA algorithms, respectively. While increasing the number of UAVs to 6, the NLoS  coverage percentages decreases to 1.86, 1.91, and 1.08.
 Also, by comparing the results of Figs. \ref{ck1} and \ref{ck2}, it can be seen that in the considered sub-urban area, by employing hybrid greedy-GA algorithm,  an NLoS coverage percentage less than 1 can be achieved with only 3 UAVs.

\subsection{UAVs' Positioning and Clustering for THz-based Network}

\begin{figure}
	\begin{center}
		\includegraphics[width=3.4 in]{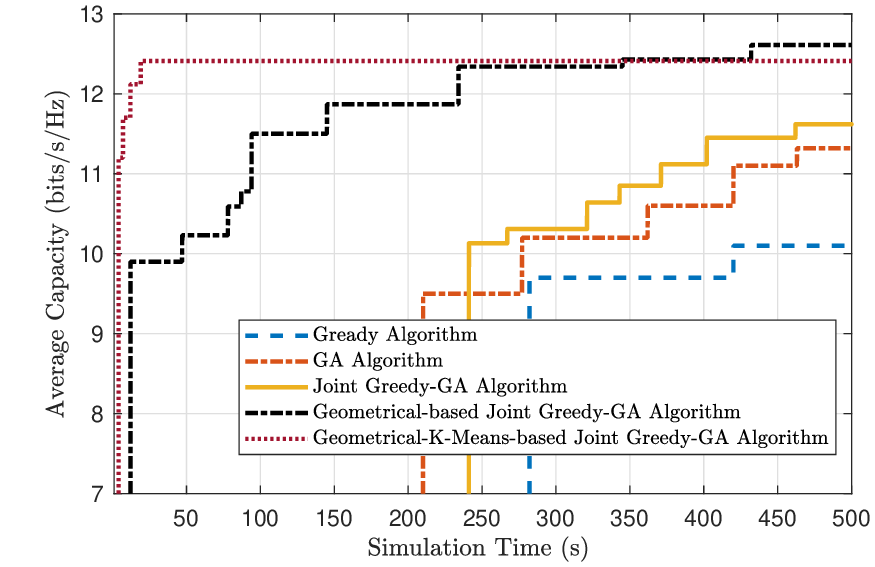}
		\caption{Comparing the average network capacity obtained by different algorithms versus running time when 4 UAVs are used to support the target area $S$.}
		\label{bv1}
	\end{center}
\end{figure}
%

In this section, by providing simulations, we investigate the performance of a UAV-based wireless network to provide high-speed THz links for a set of nodes distributed among 3D blocks. To this end, we randomly distribute 25 nodes in a dense urban area among obstacles as shown in Fig. \ref{cr}. Then, we want to locate the set of UAVs in 3D space at a height of 100 m in such a way that, while providibg LoS coverage for all the distributed nodes, it also maximizes the average capacity of the considered network. 
Unlike the previous case, here, we must have an optimal clustering of nodes along with the positioning.
For simulations, we also consider the THz frequency $f_c=188$ GHz, the relative humidity $r_h=20\%$, the environment temperature $T=25^o$, the air pressure $p=101325$ Pa, and transmit power of each antenna $P_t=5$ mW. Moreover, we consider the antenna gains of all antennas to be equal to $G'_{kn}=G_{nk}$ for $n\in\{1,...,N_u\}$ and $k\in\{1,...,N_g\}$.

\begin{figure*}
	\centering
	\subfloat[] {\includegraphics[width=6.4 in]{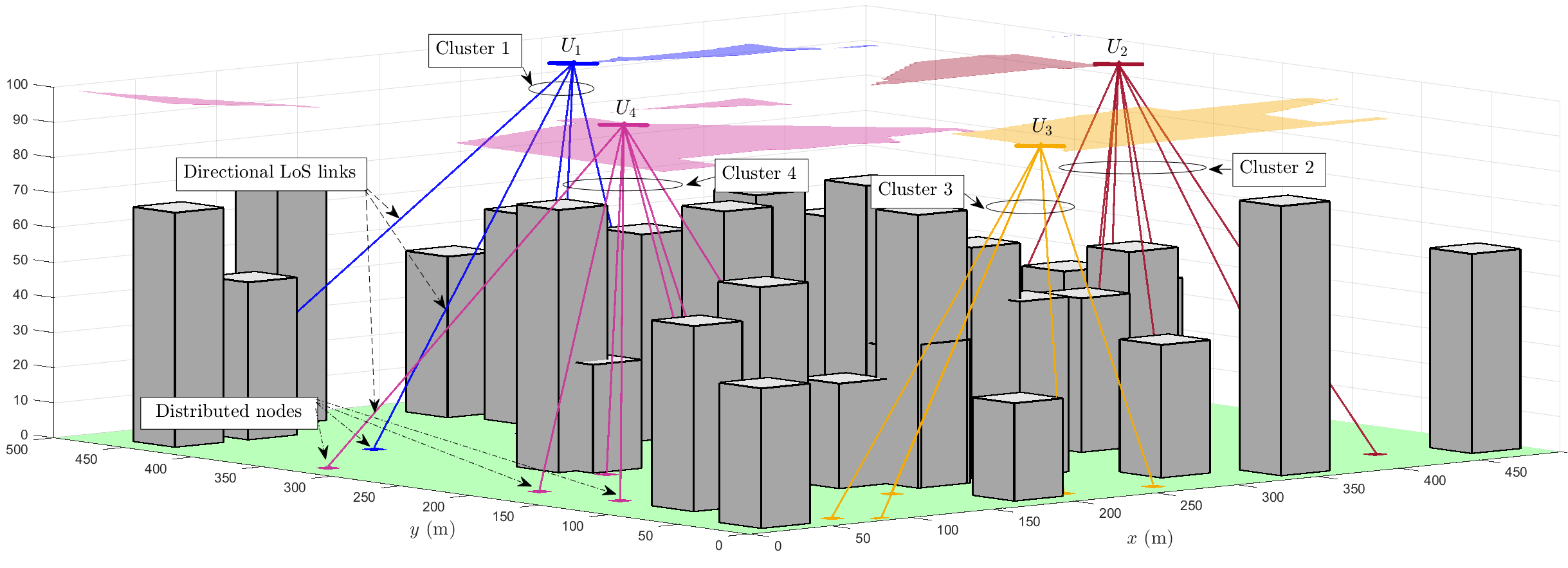}
		\label{fc1}
	}
	\hfill
	\subfloat[] {\includegraphics[width=6.4 in]{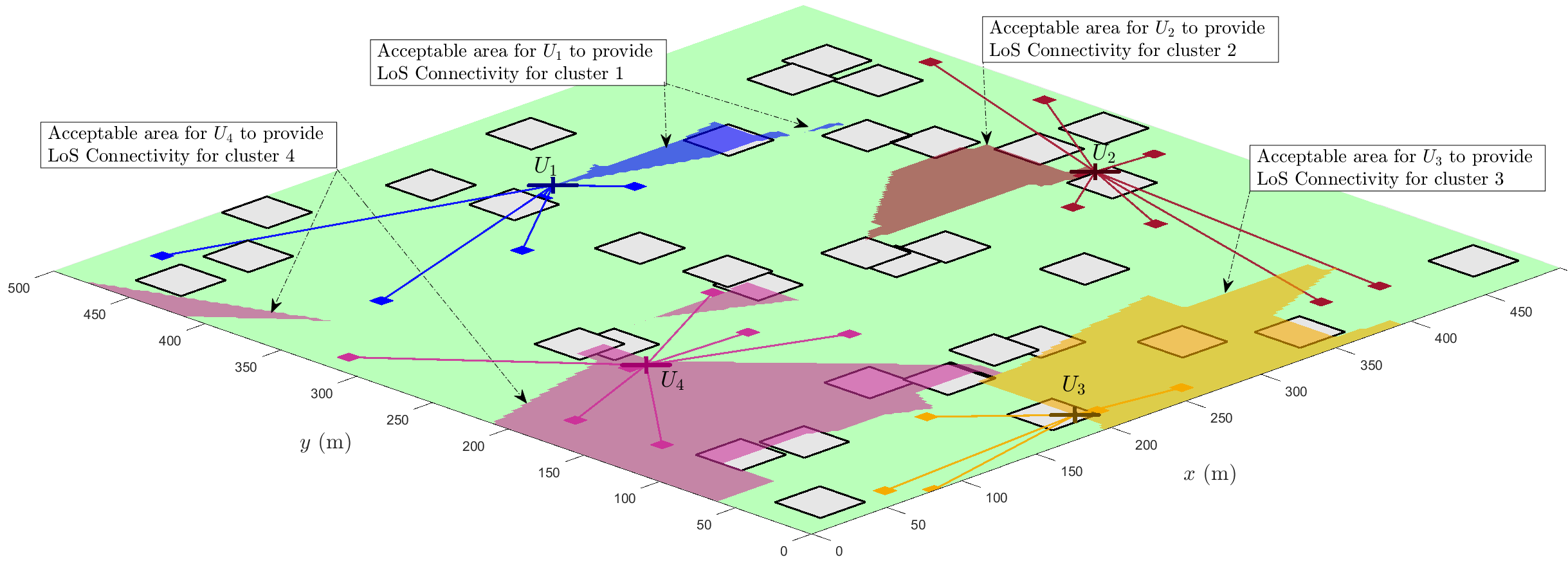}
		\label{fc2}
	}
	\caption{The 3D output of the positioning and clustering obtained by Algorithm 7 (a) 3D view, and (b) the top view of the target environment.
	}
	\label{fc}
\end{figure*}
%

Algorithms 5-7 are proposed to solve the problem of positioning and clustering of the considered system. Algorithm 5 is proposed to modify Algorithms 2-4 for the optimization problem \eqref{optim1}. In other words, to perform Algorithms 2-4 for the optimization problem \eqref{optim1}, it is necessary to run Algorithm 5 in each iterations of Algorithms 2-4 according to the explanations given. To compare the speed of convergence of these algorithms, the average capacity plotted versus time in Fig. \ref{bv1} when four UAVs are used to support the target area $S$. 
As can be seen, the modified Algorithms 2-4, which represent greedy, GA, and joint greedy-GA algorithms, respectively, do not perform well, while these algorithms performed well for the optimization problem \eqref{o1}.
The reason is that to start, these algorithms get stuck in line 9 of Algorithm 5 for many start iterations bacuse it cannot have a proper clustering of the distributed nodes. In Algorithm 6, the geometric features of the 3D blocks are used to solve the clustering problem. 
Like Algorithm 5, Algorithm 6 must be runed with one of Algorithms 2-4. Since the joint greedy-genetic algorithm has a better performance than Algorithms 2 and 3, in Fig. \ref{bv1}, Algorithm 6 is performed with Algorithm 4, and its output is denoted as geometrical-based joint greedy-GA algorithm. As can be seen, this algorithm converges significantly faster towards the optimal position and clustering of UAVs.
In the following, we also proposed a geometrical-K-means-based joint greedy-GA method in Algorithm 7, which, although it may not reach the optimal solution, it converges faster to a sub-optimal solution close to the optimal solution.
The results of the figure clearly show that Algorithm 7 achieves an average capacity of 12.4 only after 19 s.

Notice that in order to achieve the maximum capacity, it is necessary to have optimal clustering of nodes and optimal positioning of UAVs. To get a better view, in Fig. \ref{fc}, the 3D output of the positioning and clustering performed by Algorithm 7 is provided. In fact, Fig. \ref{fc} is the final 3D output of the results of Fig. \ref{bv1} for Algorithm 7. Fig. \ref{fc1} shows the 3D simulation of the target environment and Fig. \ref{fc2} is the top view of the target environment. As you can see, Algorithm 7 first performs a clustering of 25 randomly distributed nodes among the 3D obstacles of the target environment. Then, it obtains the acceptable areas $\mathcal{A}_n$s for clustering, which can be observed in Fig. \ref{fc2}. At the end, Algorithm 7  finds the sub-optimal position of the UAVs to reach the near-maximum average capacity. Based on the results of Fig. \ref{bv1}, Algorithm 7 obtains an average capacity of 12.4 only after 19 s of processing. While Algorithm 6, although it achieves a greater capacity of 12.6, it needs more than 400 s. Therefore, the output of Algorithm 7 can be used to optimize environments whose topology is changing faster (for example, nodes are mobile), because it is necessary to continuously update the clustering and positioning problem.

\begin{figure*}
	\centering
	\subfloat[] {\includegraphics[width=1.72 in]{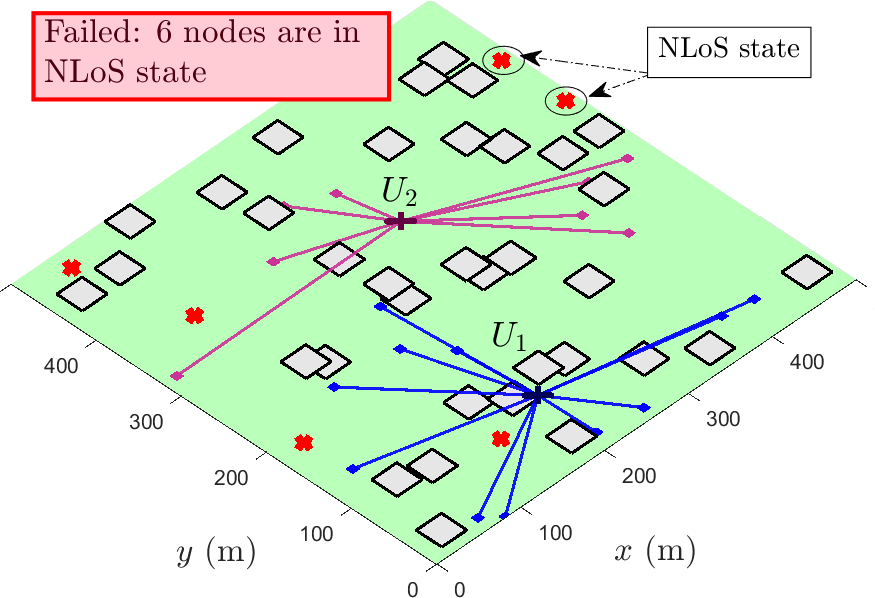}
		\label{fk1}
	}
	\hfill
	\subfloat[] {\includegraphics[width=1.72 in]{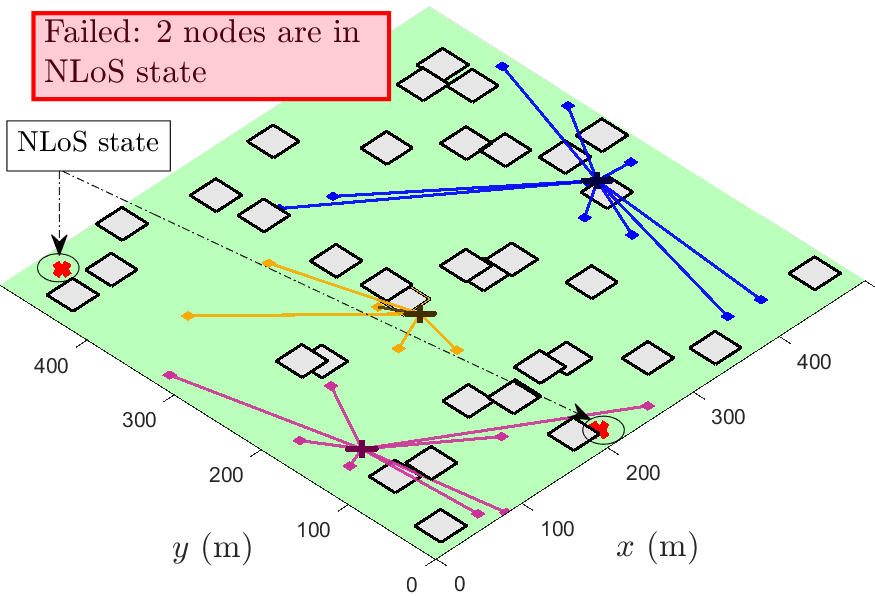}
		\label{fk2}
	}
\subfloat[] {\includegraphics[width=1.72 in]{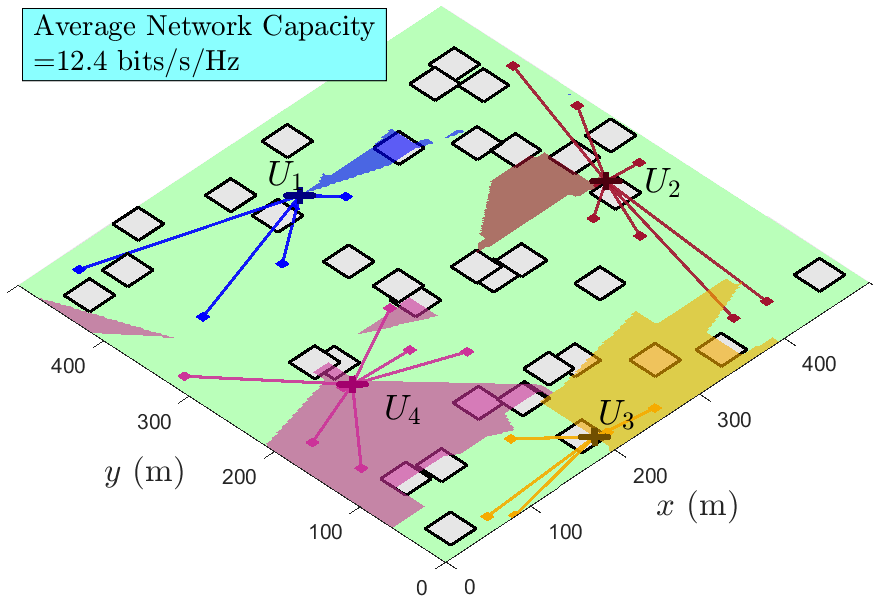}
	\label{fk3}
}
\subfloat[] {\includegraphics[width=1.72 in]{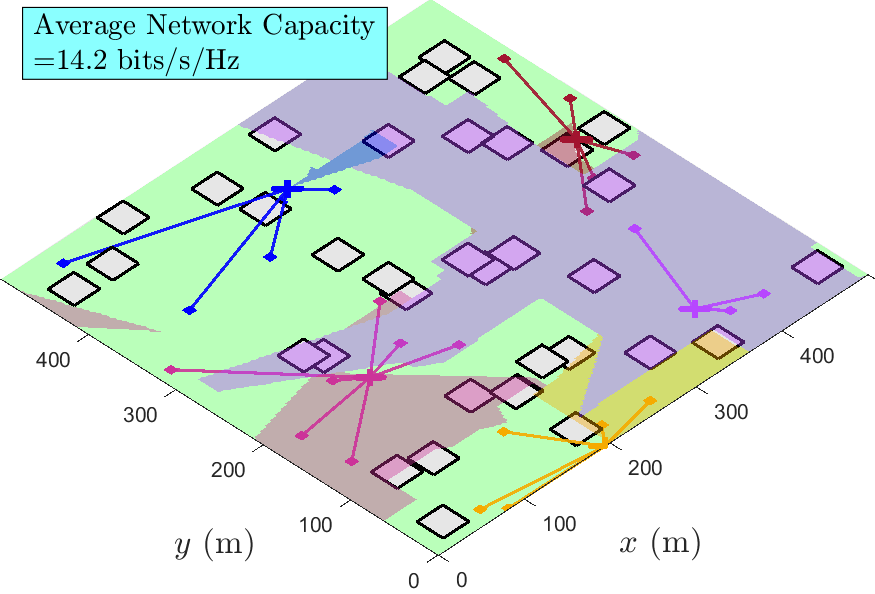}
	\label{fk4}
}
	\caption{The top view output of the positioning and clustering obtained by Algorithm 7 when the target environment environment is covered by (a) two UAVs, (b) three UAVs, (c) four UAVs, and (d) five UAVs.
	}
	\label{fk}
\end{figure*}
%

The optimization problem of \eqref{optim1} is a functions of the number of UAVs. A key remained question regarding the optimization problem \eqref{optim1} is that, for a given 3D environment, what is the minimum number of UAVs and what effect does the increase in the number of UAVs have on the average network capacity? To answer this question, the results of Fig. \ref{fk} are provided. The positioning and clustering obtained form Algorithm 7 are provided in Figs. \ref{fk1}, \ref{fk2}, \ref{fk3}, and \ref{fk4} for 2, 3, 4, and 5 UAVs, respectively. Based on the obtained results, for the considered 3D environment, we cannot create LoS coverage for all 25 randomly distributed nodes. However, with four UAVs, we achieve a capacity of 12.4 bits/s/Hz,  and with five UAVs, and we achieve an average capacity of 14.2 bits/s/Hz. In other words, with the increase in the number of UAVs, the linklength of terahertz links becomes shorter and the average capacity of the network increases.

\section{Concluding Remarks and Future Directions}
In UAV-based networks, most of the UAV's positioning analyses in the literature are based on LoS probability. However, we showed that the analysis based on LoS probability is not suitable for UAV-based high-frequency THz and optical links. In order to fill this gap and move towards realistic analyses, in this paper, we investigated the problem of UAV positioning in a realistic environment with 3D obstacles. To this end, instead of LoS probability, we studied another, more practical, concept called LoS coverage percentage. Moreover, using the obtained results, we examined a realistic optimization problem in the form of how to locate UAVs in the sky that can create LoS coverage for randomly distributed nodes among 3D obstacles. Several optimization algorithms were proposed to solve the positioning problem, which mainly provide optimal positioning of UAVs by analyzing the characteristics of 3D obstacles. Finally, by providing various 3D simulations, the optimization algorithms were evaluated in terms of performance and computational complexity.

Although there are still many open issues in the field of designing UAV-based THz networks, the results of this paper provide a step forward for the design and 3D visualization of these networks, which can be used as a preliminary step for future research directions.



\begin{thebibliography}{10} 	\balance
	\providecommand{\url}[1]{#1}
	\csname url@samestyle\endcsname
	\providecommand{\newblock}{\relax}
	\providecommand{\bibinfo}[2]{#2}
	\providecommand{\BIBentrySTDinterwordspacing}{\spaceskip=0pt\relax}
	\providecommand{\BIBentryALTinterwordstretchfactor}{4}
	\providecommand{\BIBentryALTinterwordspacing}{\spaceskip=\fontdimen2\font plus
		\BIBentryALTinterwordstretchfactor\fontdimen3\font minus
		\fontdimen4\font\relax}
	\providecommand{\BIBforeignlanguage}[2]{{%
			\expandafter\ifx\csname l@#1\endcsname\relax
			\typeout{** WARNING: IEEEtran.bst: No hyphenation pattern has been}%
			\typeout{** loaded for the language `#1'. Using the pattern for}%
			\typeout{** the default language instead.}%
			\else
			\language=\csname l@#1\endcsname
			\fi
			#2}}
	\providecommand{\BIBdecl}{\relax}
	\BIBdecl

	
	\bibitem{dabiri2022modulating}
	M.~T. Dabiri, M.~Rezaee, L.~Mohammadi, F.~Javaherian, V.~Yazdanian, M.~O.
	Hasna, and M.~Uysal, ``{Modulating retroreflector based free space optical
		link for UAV-to-ground communications},'' \emph{IEEE Transactions on Wireless
		Communications}, vol.~21, no.~10, pp. 8631--8645, 2022.
	
	\bibitem{8247211}
	Q.~Wu, Y.~Zeng, and R.~Zhang, ``Joint trajectory and communication design for
	multi-{UAV} enabled wireless networks,'' \emph{IEEE Transactions on Wireless
		Communications}, vol.~17, no.~3, pp. 2109--2121, 2018.
	
	\bibitem{8918497}
	Y.~Zeng, Q.~Wu, and R.~Zhang, ``Accessing from the sky: A tutorial on {UAV}
	communications for {5G} and beyond,'' \emph{Proceedings of the IEEE}, vol.
	107, no.~12, pp. 2327--2375, 2019.
	
	\bibitem{9457160}
	Y.~Hu, M.~Chen, W.~Saad, H.~V. Poor, and S.~Cui, ``Distributed multi-agent meta
	learning for trajectory design in wireless drone networks,'' \emph{IEEE
		Journal on Selected Areas in Communications}, vol.~39, no.~10, pp.
	3177--3192, 2021.
	
	\bibitem{9800925}
	J.~Sabzehali, V.~K. Shah, Q.~Fan, B.~Choudhury, L.~Liu, and J.~H. Reed,
	``Optimizing number, placement, and backhaul connectivity of multi-{UAV}
	networks,'' \emph{IEEE Internet of Things Journal}, vol.~9, no.~21, pp.
	21\,548--21\,560, 2022.
	
	\bibitem{gemmi2022cost}
	G.~Gemmi, R.~L. Cigno, and L.~Maccari, ``On cost-effective, reliable coverage
	for los communications in urban areas,'' \emph{IEEE Transactions on Network
		and Service Management}, vol.~19, no.~3, pp. 2767--2779, 2022.
	
	\bibitem{8478112}
	M.~T. Dabiri, S.~M.~S. Sadough, and M.~A. Khalighi, ``Channel modeling and
	parameter optimization for hovering {UAV}-based free-space optical links,''
	\emph{IEEE Journal on Selected Areas in Communications}, vol.~36, no.~9, pp.
	2104--2113, 2018.
	
	\bibitem{gemmi2022properties}
	G.~Gemmi, R.~L. Cigno, and L.~Maccari, ``On the properties of next generation
	wireless backhaul,'' \emph{IEEE Transactions on Network Science and
		Engineering}, vol.~10, no.~1, pp. 166--177, 2022.
	
	\bibitem{hu2020low}
	Q.~Hu, Y.~Cai, A.~Liu, G.~Yu, and G.~Y. Li, ``{Low-complexity joint resource
		allocation and trajectory design for UAV-aided relay networks with the
		segmented ray-tracing channel model},'' \emph{IEEE Transactions on Wireless
		Communications}, vol.~19, no.~9, pp. 6179--6195, 2020.
	
	\bibitem{9709500}
	X.~Xia, Y.~Wang, K.~Xu, and Y.~Xu, ``Toward digitalizing the wireless
	environment: A unified {A2G} information and energy delivery framework based
	on binary channel feature map,'' \emph{IEEE Transactions on Wireless
		Communications}, vol.~21, no.~8, pp. 6448--6463, 2022.
	
	\bibitem{zeng2021toward}
	Y.~Zeng and X.~Xu, ``{Toward environment-aware 6G communications via channel
		knowledge map},'' \emph{IEEE Wireless Communications}, vol.~28, no.~3, pp.
	84--91, 2021.
	
	\bibitem{zeng2021simultaneous}
	Y.~Zeng, X.~Xu, S.~Jin, and R.~Zhang, ``{Simultaneous navigation and radio
		mapping for cellular-connected UAV with deep reinforcement learning},''
	\emph{IEEE Transactions on Wireless Communications}, vol.~20, no.~7, pp.
	4205--4220, 2021.
	
	\bibitem{9200666}
	M.~T. Dabiri, M.~Rezaee, V.~Yazdanian, B.~Maham, W.~Saad, and C.~S. Hong,
	``{3D} channel characterization and performance analysis of {UAV}-assisted
	millimeter wave links,'' \emph{IEEE Transactions on Wireless Communications},
	vol.~20, no.~1, pp. 110--125, 2021.
	
	\bibitem{wang2020placement}
	W.~Wang, H.~Dai, C.~Dong, X.~Cheng, X.~Wang, P.~Yang, G.~Chen, and W.~Dou,
	``{Placement of unmanned aerial vehicles for directional coverage in 3D
		space},'' \emph{IEEE/ACM transactions on networking}, vol.~28, no.~2, pp.
	888--901, 2020.
	
	\bibitem{sabzehali20213d}
	J.~Sabzehali, V.~K. Shah, H.~S. Dhillon, and J.~H. Reed, ``{3D placement and
		orientation of mmWave-based UAVs for guaranteed LoS coverage},'' \emph{IEEE
		Wireless Communications Letters}, vol.~10, no.~8, pp. 1662--1666, 2021.
	
	\bibitem{9044827}
	J.~Zhao, J.~Liu, J.~Jiang, and F.~Gao, ``Efficient deployment with geometric
	analysis for {mmWave UAV} communications,'' \emph{IEEE Wireless
		Communications Letters}, vol.~9, no.~7, pp. 1115--1119, 2020.
	
	\bibitem{lin2021adaptive}
	N.~Lin, Y.~Liu, L.~Zhao, D.~O. Wu, and Y.~Wang, ``{An adaptive UAV deployment
		scheme for emergency networking},'' \emph{IEEE Transactions on Wireless
		Communications}, vol.~21, no.~4, pp. 2383--2398, 2021.
	
	\bibitem{li2022geometric}
	F.~Li, C.~He, X.~Li, J.~Peng, and K.~Yang, ``{Geometric analysis-based 3D
		anti-block UAV deployment for mmWave communications},'' \emph{IEEE
		Communications Letters}, vol.~26, no.~11, pp. 2799--2803, 2022.
	
	\bibitem{yi2022joint}
	P.~Yi, L.~Zhu, L.~Zhu, Z.~Xiao, Z.~Han, and X.-G. Xia, ``{Joint 3-D positioning
		and power allocation for UAV relay aided by geographic information},''
	\emph{IEEE Transactions on Wireless Communications}, vol.~21, no.~10, pp.
	8148--8162, 2022.
	
	\bibitem{yi20233}
	P.~Yi, L.~Zhu, Z.~Xiao, R.~Zhang, Z.~Han, and X.-G. Xia, ``{3-D Positioning and
		Resource Allocation for Multi-UAV Base Stations Under Blockage-Aware Channel
		Model},'' \emph{IEEE Transactions on Wireless Communications}, 2023.
	
	\bibitem{tang2021performance}
	W.~Tang, H.~Zhang, and Y.~He, ``{Performance analysis of power control in urban
		UAV networks with 3D blockage effects},'' \emph{IEEE Transactions on
		Vehicular Technology}, vol.~71, no.~1, pp. 626--638, 2021.
	
	\bibitem{zhu2022geometry}
	Q.~Zhu, F.~Bai, M.~Pang, J.~Li, W.~Zhong, X.~Chen, and K.~Mao,
	``{Geometry-based stochastic line-of-sight probability model for A2G channels
		under urban scenarios},'' \emph{IEEE Transactions on Antennas and
		Propagation}, vol.~70, no.~7, pp. 5784--5794, 2022.
	
	\bibitem{dOHA_WE}
	\BIBentryALTinterwordspacing
	O.~Buildings. Qatar, doha, west bay towers. [Online]. Available:
	\url{https://osmbuildings.org/}
	\BIBentrySTDinterwordspacing
	
	\bibitem{balanis2016antenna}
	C.~A. Balanis, \emph{Antenna theory: analysis and design}.\hskip 1em plus 0.5em
	minus 0.4em\relax John wiley \& sons, 2016.
	
	\bibitem{3gppf}
	3GPP, ``{Study on channel model for frequencies from 0.5 to 100 GHz (Release
		14)},'' \emph{3GPP TR 38.901 V14.1.1}, Jul. 2017.
	
	\bibitem{9998554}
	M.~T. Dabiri, M.~Hasna, and W.~Saad, ``Downlink interference analysis of
	{UAV-based mmWave} fronthaul for small cell networks,'' \emph{IEEE
		Transactions on Vehicular Technology}, vol.~72, no.~5, pp. 5560--5575, 2023.
	
	\bibitem{boulogeorgos2019analytical}
	A.-A.~A. Boulogeorgos, E.~N. Papasotiriou, and A.~Alexiou, ``{Analytical
		performance assessment of THz wireless systems},'' \emph{IEEE Access},
	vol.~7, pp. 11\,436--11\,453, 2019.
	
	\bibitem{kokkoniemi2021line}
	J.~Kokkoniemi, J.~Lehtom{\"a}ki, and M.~Juntti, ``A line-of-sight channel model
	for the 100--450 gigahertz frequency band,'' \emph{EURASIP Journal on
		Wireless Communications and Networking}, vol. 2021, no.~1, pp. 1--15, 2021.
	
	\bibitem{alduchov1996improved}
	O.~A. Alduchov and R.~E. Eskridge, ``Improved magnus form approximation of
	saturation vapor pressure,'' \emph{Journal of Applied Meteorology and
		Climatology}, vol.~35, no.~4, pp. 601--609, 1996.
	
	\bibitem{likas2003global}
	A.~Likas, N.~Vlassis, and J.~J. Verbeek, ``The global k-means clustering
	algorithm,'' \emph{Pattern recognition}, vol.~36, no.~2, pp. 451--461, 2003.
	
\end{thebibliography}
\end{document}